\newcommand{\Keff}{K_\mathrm{eff}}
\newcommand{\Katt}{K_\mathrm{att}}

\newcommand{\Lmean}{\langle l_\parallel \rangle}
\newcommand{\Lproj}{l_\parallel}
\newcommand{\amean}{\langle \alpha \rangle}
\newcommand{\aangle}{\alpha}
\newcommand{\lchar}{l_\mathrm{char}}
\newcommand{\Radius}{R}
\newcommand{\SurfAngle}{\theta}
\newcommand{\lmin}{l_\mathrm{min}}
\newcommand{\tmin}{\tau_\mathrm{min}}
\newcommand{\latt}{l_\mathrm{att}}
\newcommand{\length}{l}
\newcommand{\plength}{p_\mathrm{l}}
\newcommand{\patt}{p_\mathrm{att}}
\newcommand{\Pattmin}{P_\mathrm{att,min}}
\newcommand{\Pattangle}{P_\mathrm{att,\theta}}
\newcommand{\velocity}{v}
\newcommand{\Pgamma}{\Gamma_\mathrm{0}}
\newcommand{\surfdens}{\rho}
\newcommand{\Natt}{N_\mathrm{att}}
\newcommand{\Ntot}{N_\mathrm{tot}}
\newcommand{\Patt}{\Gamma}
\newcommand{\AttTime}{T}
\newcommand{\Keffatt}{K_\mathrm{eff,att}}
\newcommand{\lturn}{l_\mathrm{tu}}
\newcommand{\pturn}{p_\mathrm{tu}}
\newcommand{\Pturn}{P_\mathrm{tu}}
\newcommand{\ptwo}{p_\mathrm{2}}
\newcommand{\ptotallength}{p_\mathrm{L}}
\newcommand{\lzero}{l_\mathrm{0}}
\newcommand{\height}{h}
\newcommand{\tcontact}{\tau_\mathrm{con}}
\newcommand{\surface}{S}
\newcommand{\Ntotal}{N}
\newcommand{\Nleft}{N_\mathrm{l}}
\newcommand{\Nright}{N_\mathrm{r}}
\newcommand{\Fleft}{F_\mathrm{l}}
\newcommand{\Fright}{F_\mathrm{r}}
\newcommand{\Fleftsteady}{F_\mathrm{l}^\mathrm{(st)}}
\newcommand{\Frightsteady}{F_\mathrm{r}^\mathrm{(st)}}
\newcommand{\Nfree}{N_\mathrm{free}}
\newcommand{\Plr}{P}
\newcommand{\TransitionMatrix}{T}

\newcommand{\Forcenorm}{F_\perp}
\newcommand{\Forcepara}{F_\parallel}

\newcommand{\kattlcell}{k_\mathrm{att,l}^\mathrm{(cell)}}
\newcommand{\kattrcell}{k_\mathrm{att,r}^\mathrm{(cell)}}
\newcommand{\kattlcolony}{k_\mathrm{att,l}^\mathrm{(col)}}
\newcommand{\kattrcolony}{k_\mathrm{att,r}^\mathrm{(col)}}
\newcommand{\kdetl}{k_\mathrm{det,l}}
\newcommand{\kdetr}{k_\mathrm{det,r}}
\newcommand{\kmotl}{k_\mathrm{mot,l}}
\newcommand{\kmotr}{k_\mathrm{mot,r}}
\newcommand{\Fstall}{F_\mathrm{s}}
\newcommand{\mobtrans}{\mu_\mathrm{t}}
\newcommand{\vret}{v_\mathrm{ret}}
\newcommand{\Kdet}{K_\mathrm{det}}
\newcommand{\vcell}{v_\mathrm{c}}
\newcommand{\lc}{l_\mathrm{c}}
\newcommand{\vc}{v_\mathrm{c}}
\newcommand{\tc}{\tau_\mathrm{c}}

\newcommand{\ria}{\mathbf{r}_i^\mathrm{(a)}}
\newcommand{\rib}{\mathbf{r}_i^\mathrm{(b)}}

\newcommand{\rcom}{\mathbf{r}_i^\mathrm{(com)}}
\newcommand{\dcocci}{d_\mathrm{cocci}}
\newcommand{\vpro}{v_\mathrm{pro}}

\newcommand{\dt}{\Delta t}

\newcommand{\lcontour}{l_k^\mathrm{(cont)}}
\newcommand{\lfree}{l_k^\mathrm{(free)}}

\newcommand{\dover}{\Delta d_\mathrm{ov}}

\newcommand{\kcs}{k_\mathrm{cs}}
\newcommand{\mobrotat}{\mu_\mathrm{rotat}}
\newcommand{\kspring}{k_\mathrm{pili}}

\newcommand{\Fdetpsone}{F^\mathrm{(1)}_\mathrm{det}}
\newcommand{\Fdetpstwo}{F^\mathrm{(2)}_\mathrm{det}}
\newcommand{\tdetpsone}{\tau^\mathrm{(1)}_\mathrm{det}}
\newcommand{\tdetpstwo}{\tau^\mathrm{(2)}_\mathrm{det}}

\newcommand{\Ftot}{\mathbf{F}_i^\mathrm{(tot)}} 
\newcommand{\Ttot}{\mathbf{T}_i^\mathrm{(tot)}} 
\newcommand{\wtot}{\mathbf{w}_i^\mathrm{(tot)}} 
\newcommand{\Fjsub}{\mathbf{F}_j^\mathrm{(cs)}} 
\newcommand{\djsub}{\mathbf{d}_j^\mathrm{(cs)}} 
\newcommand{\Fkps}{\mathbf{F}_k^\mathrm{(ps)}} 
\newcommand{\dkps}{\mathbf{d}_k^\mathrm{(ps)}} 
\newcommand{\dkpp}{\mathbf{d}_k^\mathrm{(pp)}} 
\newcommand{\Finorm}{F_{i}^{(\mathrm{norm})}} 
\newcommand{\slidcoeff}{\beta} 
\newcommand{\lallxyz}{l_\mathrm{xyz}^\mathrm{(all)}}
\newcommand{\lattxy}{l_\mathrm{xy}^\mathrm{(att)}}
\newcommand{\viscosity}{\eta}
\newcommand{\Fleftperp}{F_\mathrm{l}^\perp}
\newcommand{\Frightperp}{F_\mathrm{r}^\perp}
\newcommand{\Fleftpara}{F_\mathrm{l}^\parallel}
\newcommand{\Frightpara}{F_\mathrm{r}^\parallel}
\newcommand{\Lleftperp}{L_\mathrm{l}^\perp}
\newcommand{\Lrightperp}{L_\mathrm{r}^\perp}
\newcommand{\Lleftpara}{L_\mathrm{l}^\parallel}
\newcommand{\Lrightpara}{L_\mathrm{r}^\parallel}

\newcommand{\lleft}{l_\mathrm{l}}
\newcommand{\lleftdiff}{\dot{l}_\mathrm{l}}
\newcommand{\Lright}{L_\mathrm{r}}
\newcommand{\Lrightdiff}{\dot{L}_\mathrm{r}}
\newcommand{\lright}{l_\mathrm{r}}
\newcommand{\lrightdiff}{\dot{l}_\mathrm{r}}
\newcommand{\Fleftdiff}{\dot{F}_\mathrm{l}}
\newcommand{\Frightdiff}{\dot{F}_\mathrm{r}}
\newcommand{\Ftotlr}{F_\mathrm{tot}}

\newcommand{\ood}{\Delta \phi}
\newcommand{\oo}{\phi_i}

\newcommand{\Lleft}{L_\mathrm{l}}
\newcommand{\Lleftdiff}{\dot{L}_\mathrm{l}}
\newcommand{\Lleftdiffpara}{\dot{L}_\mathrm{l}^\parallel}
\newcommand{\Lrightdiffpara}{\dot{L}_\mathrm{r}^\parallel}

\documentclass[%
reprint,
amsmath,amssymb,
aps,
]{revtex4-1}

\usepackage{graphicx}
\usepackage{dcolumn}
\usepackage{bm}
\usepackage{xcolor}

\begin{document}
	
	
	\title{How bacterial cells and colonies move on solid substrates}
	
	\author{Wolfram P\"onisch}
	\affiliation{Max Planck Institute for the Physics of Complex Systems, 01187 Dresden, Germany}
	\affiliation{MRC Laboratory for Molecular Cell Biology, University College London, London, WC1E 6BT, UK}
	\author{Christoph A. Weber}
	\affiliation{Paulson School of Engineering and Applied Sciences, Harvard University, Cambridge, MA 02138, USA}
	\affiliation{Max Planck Institute for the Physics of Complex Systems, 01187 Dresden, Germany}

	
	\author{Vasily Zaburdaev}
	\email{vasily.zaburdaev@fau.de}
	\affiliation{Max Planck Institute for the Physics of Complex Systems, 01187 Dresden, Germany}
	\affiliation{Institute of Supercomputing Technologies, Lobachevsky State University of Nizhny Novgorod, Nizhny Novgorod 603140, Russia}
	\affiliation{Friedrich-Alexander-Universit\"at Erlangen-N\"urnberg, 91058 Erlangen, Germany}
	
	%
	%
	%
	
	\date{\today}
	
	\begin{abstract}
		Many bacteria rely on active cell appendages, such as type IV pili, to move over substrates and interact with neighboring cells. 
		Here, we study the motion of individual cells and bacterial colonies, mediated by the collective interactions of multiple pili.
		It was shown experimentally that the substrate motility of \textit{Neisseria gonorrhoeae} cells can be described as a persistent random walk with a persistence length that exceeds the mean pili length. Moreover, the persistence length increases for a higher number of pili per cell. 
		With the help of a simple, tractable stochastic model, we test whether a tug-of-war without directional memory can explain the persistent motion of single \textit{Neisseria gonorrhoeae} cells.
		While persistent motion of single cells indeed emerges naturally in the model, a tug-of-war alone is not capable of explaining the motility of microcolonies, which becomes weaker with increasing colony size.
		We suggest sliding friction between the microcolonies and the substrate as the missing ingredient. While such friction almost does not affect the general mechanism of single cell motility, it has a strong effect on colony motility. 
		We validate the theoretical predictions by using a three-dimensional computational model that includes explicit details of the pili dynamics, force generation and geometry of cells.
	\end{abstract}
	
	\maketitle
	
	
	\section{\label{sec:Introduction} Introduction}
\begin{figure}[tb]
		\includegraphics{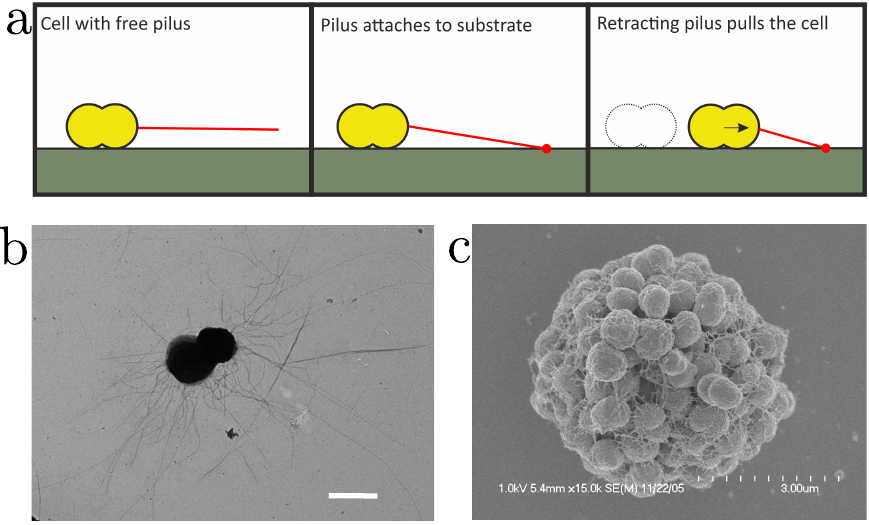}%
		\caption{\label{figmechanism} The mechanism of pili-mediated motion of aggregates. (a) Grappling hook mechanism of type IV pili. Pili are initially not attached to a substrate. When a retracting pilus is attached to a substrate, it will generate a pulling force that can mediate motion of the cell. (b) Electron microscope image of a single cell of \textit{N. gonorrhoeae} with several individual pili. Scale bar~${=1\ \mu\mathrm{m}}$. (c) Electron microscope image of a microcolony of \textit{N. gonorrhoeae} sitting on a substrate. The electron microscope images were contributed by Prof. Nicolas Biais (Brooklyn College, CUNY, NY, USA). }
	\end{figure}

	When one thinks of moving bacteria, one typically imagines cells that swim with the help of flagella~\cite{watnick:2000, berg:2008}. However, most bacteria exist in close association with substrates and developed a variety of tools to mediate motion over a wide range of surfaces such as hulls of ships~\cite{townsin:2003}, medical catheters~\cite{trautner:2004} and within the human body~\cite{macfarlane:1997, zijnge:2010}. Many bacteria, both gram-negative (such as \textit{Neisseria meningitidis}~\cite{imhaus:2014}, \textit{Pseudomonas aeruginosa}~\cite{gibiansky:2010, conrad:2011, brill:2017} and \textit{Myxococcus xanthus}~\cite{mattick:2002}) and gram-positive (such as \textit{Clostridium difficile}~\cite{pelicic:2008, piepenbrink:2016}), use so-called type IV pili for the motion over substrates and attachment to other cells.
	Pili are several microns long filaments that emerge from the surface of cells. Due to cycles of protrusion, attachment, and retraction, pili can mediate motion of the cell by a mechanism reminiscent of a grappling hook~\cite{merz:2002} (see Fig.~\ref{figmechanism}a). During this process, individual retracting pili can generate forces of up to ${100-180\ \mathrm{pN}}$~\cite{maier:2002, biais:2010, marathe:2014}. The retraction is caused by the molecular motor pilT via disassembly of the pilus filament~\cite{merz:20002, maier:2015}.
	Additionally, pili mediate the formation of bacterial microcolonies, precursors of biofilms, by pili-pili-interactions. Bacteria without pili are no longer able to agglomerate~\cite{taktikos:2015, weber:2015, oldewurtel:2015, poenisch:2017}.
	
	While in this study we focus on the bacterium \textit{N. gonorrhoeae} (see Fig.~\ref{figmechanism}b), the suggested models and results can be broadly applied to other bacteria that use type IV pili for the motion on a substrate. For the motility of \textit{N. gonorrhoeae} a wide range of experimental observations have been described~\cite{holz:2010, zaburdaev:2014, marathe:2014, eriksson:2015, taktikos:2015}, making it a suitable model organism for our study. 
	One of the interesting observations is that the motion of individual bacteria can be described as a persistent random walk with a characteristic length scale that exceeds the mean pili length~\cite{holz:2010}. 
	Previously, it has been suggested that the persistence of the motion originates from a directional memory of the pili~\cite{marathe:2014}. The directional memory is generated by re-elongation of fully retracted pili in the same direction and by the bundling of neighboring pili. 
	In the same study, it has been suggested that the directional memory is also responsible for the experimentally observed increasing persistence length of the motility for cells with an increasing number of pili~\cite{holz:2010, marathe:2014}.
	
	Next to single cell motility, pili are also involved in the formation of bacterial microcolonies, mediated by pili-pili-interactions~\cite{weber:2015, taktikos:2015}. The microcolonies possess a spherical shape (see Fig.~\ref{figmechanism}c) and can move on substrates, similarly to single cells. By analyzing trajectories of microcolonies moving over a substrate it has been shown that the microcolonies exhibit a decreasing motility with increasing colony size~\cite{taktikos:2015}. 
	
	By imagining microcolonies as single cells with increasing size and number of pili, we are confronted with a discrepancy: while a higher number of pili enhances the motility of single cells, the motility of microcolonies is reduced. One possible reason for this behavior could be the increasing hydrodynamic friction of microcolonies. Both, single cells and large colonies, moving on a substrate within a fluid are subject to a hydrodynamic drag force. Neglecting the impact of the substrate on hydrodynamic flows, the Einstein-Stokes equation provides an approximative estimate for the drag force, ${F = 6 \pi \viscosity \Radius \velocity}$, with the viscosity $\viscosity$ of the surrounding fluid. Cells and colonies move with a characteristic velocity of around ${2\ \mu \mathrm{m} \slash \mathrm{s}}$ over a substrate, corresponding to a drag force of less than ${0.2\ \mathrm{pN}}$. If a cell is approximated as a sphere moving parallel to a wall in a viscous fluid, the resulting hydrodynamic effects of the substrate increase the drag force and increase the friction force up to three to four times~\cite{goldman:1967}. These forces are considerably smaller than the force generated by an individual pilus, which can reach up to ${100 - 180\ \mathrm{pN}}$~\cite{maier:2015}, and smaller than the lowest measured unbinding forces with the substrates of  ${1.2\ \mathrm{pN}}$~\cite{marathe:2014}. 
	If the hydrodynamic friction only weakly affects the motility of cells and microcolonies, then what can explain the decreasing motility of colonies with increasing colony size? Here, we suggest that sliding friction between the substrate and the colony can be the key to explaining the decrease in motility of microcolonies with increasing size, as seen in the experiment~\cite{taktikos:2015}.
	
	As even the motion of a single cell exhibits quite complex behavior, we developed a quasi one-dimensional model to understand the physical mechanism underlying pili-mediated cell motility. We show that a tug-of-war mechanism can be sufficient to explain the emergence of persistent motion of individual cells without invoking additional assumptions, such as directional memory. The tug-of-war mechanism also recovers the increasing persistence with increasing total number of pili of an individual cell.
	We also apply the stochastic model to explain the reduced motility of microcolonies and discuss the role of sliding friction. 
	To benchmark the results of the stochastic model, we use a previously developed three-dimensional model (see Refs.~\cite{poenisch:2017, poenisch:20172} and summarized in Appendix~\ref{sec:appendixCompModel}). It reproduces the motion of a cell by accounting for the dynamics of its pili and corresponding forces. In Appendix~\ref{sec:appendixCompModel}, we introduce an extension of the three-dimensional model, allowing us to consider the sliding friction between the substrate and cells and colonies.
	
	Before we introduce the methods that we use to describe substrate motility of cells and aggregates, we first summarize the main features of pili-mediated motion of \textit{N. gonorrhoeae} known from experiments and introduce two parameter sets, corresponding to different pictures of the pili binding and unbinding dynamics.

	\subsection{\label{sec:Properties} Properties of \textit{N. gonorrhoeae} bacteria and its pilus apparatus}
	
	Bacteria of \textit{N. gonorrhoeae} have the shape of a diplococcus, two spheres (each with a radius of around ${0.5\ \mu \mathrm{m}}$~\cite{zaburdaev:2014}) attached to each other forming a dumbbell~\cite{craig:2004} (see Fig.~\ref{figmechanism}b). A single cell possesses around 10-20 pili~\cite{zaburdaev:2014, holz:2010, marathe:2014} that are homogeneously distributed on its surface~\cite{eriksson:2015}. 
	With the help of electron microscopy, it was shown that pili have a diameter of approximately $6\ \mathrm{nm}$~\cite{mattick:2002, wang:2017} and an exponential length distribution with a mean length of ${\lchar = 0.9-1.2\ \mu \mathrm{m}}$~\cite{holz:2010,zaburdaev:2014}. Pili can be described as semiflexible polymers with a persistence length of ${5~\mu \mathrm{m}}$~\cite{skerker:2001, haan:2016}.
	Pili are assembled from pilin subunits within the membrane of the bacteria and protrude with the velocity ${\vpro \approx 2\ \mu \mathrm{m} \slash \mathrm{s}}$. They stochastically switch to the retraction state, in which a pilus retracts with a velocity ${\vret \approx 1-2\ \mu \mathrm{m} \slash \mathrm{s}}$, due to the disassembly of the filament. The molecular motor pilT is responsible for the retraction of the pilus~\cite{maier:2013, maier:2015}. The retraction continues until the pilus is completely retracted. 
	
	Pili bind stochastically to substrates. The exact mechanism of how type IV pili bind to substrates is not completely understood. It has been suggested that pili possess key residues of the pilin subunits that are involved in adhesion and are only exposed at the tip~\cite{lee:1994, wong:1995, harvey:2009}. Other studies suggest that there might be further proteins associated with pili binding that are not only located at the tip of the pili but can also be rarely found along the filament~\cite{giltner:2006, heiniger:2010}. Additionally, in a recent study, it has been suggested that pili in the protrusion state will immediately switch to the retraction state after attachment~\cite{chang:2016}. 
	
	If an attached pilus is retracting, it generates a pulling force. The sum of all pili-mediated forces causes the translational and rotational motion of the cells.
	Furthermore, a pulling force $F$ acting on a pilus affects its retraction velocity, a process that is known as stalling with the characteristic stalling force~$\Fstall$. Stalling can be described by the following force-velocity relation: 
	\begin{equation}
	v = - \max \left[ 0,\ \vret \left( 1 - \frac{F}{\Fstall}\right) \right],
	\label{eq:stalling}
	\end{equation}
	where the minus sign describes the shortening of the pilus during retraction. The stalling force is approximately ${\Fstall \approx 100 - 180\ \mathrm{pN}}$ and represents the maximal force a retracting pilus can generate~\cite{maier:2013,julicher:1999}. The rate~$\Kdet$ with which a pilus detaches from the substrate is also force-dependent~\cite{marathe:2014} and can be approximated by  
	\begin{equation}\label{eq:detrate}
	\Kdet(F) = \frac{1}{\tdetpsone \exp \left(- \frac{F}{\Fdetpsone}\right) + \tdetpstwo \exp \left(- \frac{F}{\Fdetpstwo}\right)}.
	\end{equation}
	While the general exponential form of this equation is motivated by Kramer~\cite{kramers:1940} and Bell et al.~\cite{bell:1978}, the sum of two exponential functions in the denominator is based on experimental measurements~\cite{marathe:2014}.
	
	\subsection{\label{sec:TwoPictures} Two pictures of the binding and unbinding dynamics of pili}
	
	From previous studies~\cite{marathe:2014, zaburdaev:2014, brill:2017}, we are confronted with two different regimes of how pili bind to and unbind from the substrate. These two regimes are incorporated in this study by two parameter sets for the detachment forces, $\Fdetpsone$ and $\Fdetpstwo$, and detachment times, $\tdetpsone$ and $\tdetpstwo$, referred to \textit{fast} and \textit{slow}  (see Table~\ref{table_parametersets}).
	
	\begin{table}[!tb]
		\caption{\label{table_parametersets}%
			Detachment forces, $\Fdetpsone$ and $\Fdetpstwo$, detachment times, $\tdetpsone$ and $\tdetpstwo$ and attachment rates~$\Katt$ for the \textit{fast} and \textit{slow} parameter sets. The values of the \textit{fast} parameter set were taken from~\cite{marathe:2014}.
		}
		\begin{ruledtabular}
			\begin{tabular}{lcr}
				\textrm{ }&
				\textrm{fast}&
				\textrm{slow}\\
				\colrule
				$\Fdetpsone\ [\mathrm{pN}]$  & 1.24 & 180  \\
				$\Fdetpstwo\ [\mathrm{pN}]$ & 33.8 &    \\ 
				$\tdetpsone\ [\mathrm{s}]$  & 0.85 & 10  \\
				$\tdetpstwo\ [\mathrm{s}]$ & 0.04 & 0  \\
				$\Katt\ [\mathrm{Hz}]$  & 15 & 0.5  \\
			\end{tabular}
		\end{ruledtabular}
	\end{table}
	
	The parameters of the \textit{fast} set were taken from the literature for the attachment of \textit{N. gonorrhoeae} pili to BSA coated silica beads~\cite{marathe:2014}. For this parameter set, pili only bind weakly to the substrate, but frequently. 
	For the \textit{slow} parameter set, pili bind less frequently, but more stably. This picture has been suggested in a previous study~\cite{zaburdaev:2014}.
	
	In the following, we investigate whether the collective interactions of pili of a single cell or a colony, particularly the force-dependent detachment, is sufficient to be the driving mechanism of the persistent motion of cells. 
	
	\section{\label{sec:StochasticModel} Stochastic model for pili-mediated cell and colony motion}
	
	In this section, we introduce a stochastic model for cell motility, which relies on a tug-of-war mechanism. Such mechanisms are frequently found to explain cargo transport along filaments by molecular motors~\cite{muller:2008, soppina:2009, hendricks:2010}, where molecular motors pull on a cargo in opposite directions. Due to the stochastic attachment of the motors to the filament and slip bond behavior of attached motors, complex collective behavior, such as persistent motion, can emerge.
	
	Here, we map the three-dimensional process of cell and colony motility on a two-dimensional substrate to a one-dimensional stochastic tug-of-war model. 
	To perform the mapping, we explicitly consider the geometry of the pili, operating both for a single cell and a colony, and analytically estimate the parameters required for the stochastic model.
	Afterwards, we define the master equation for the stochastic model and use it to describe the motility of individual cells and colonies.
	
	\subsection{\label{sec:Mapping} Mapping of binding and geometric features of pili to the one-dimensional motion}
	
	For simplicity, we initially focus on individual cells and later extend our description for microcolonies. This can be done because we assume an identical geometry for cells and colonies: both are described as spheres possessing multiple pili. For the mapping of the binding and geometric features of the pili, the diplococcus shape of single cells can be neglected.
	
	\begin{figure*}[tb]
		\includegraphics[width=14cm]{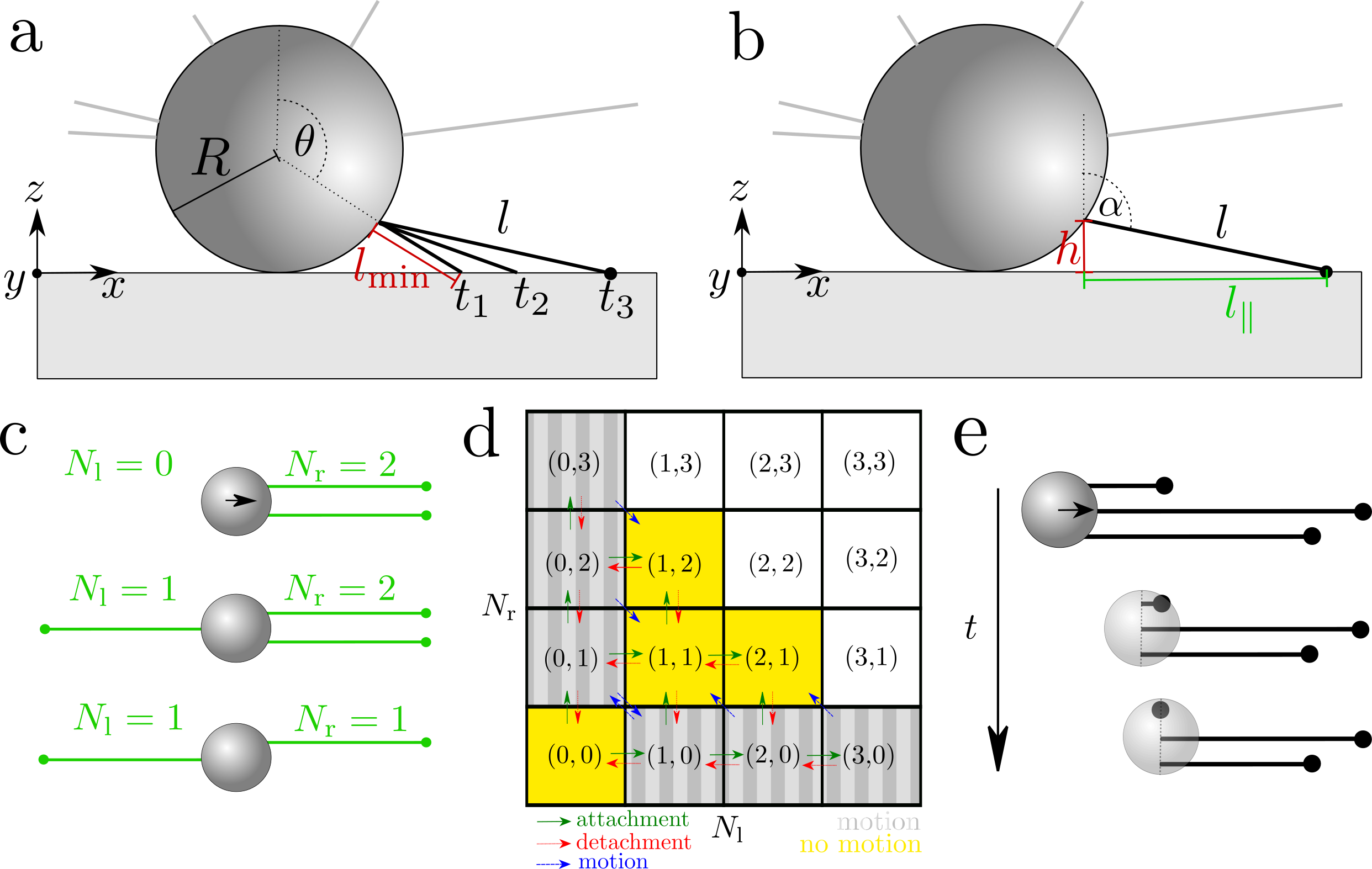}%
		\caption{\label{figstochastic} Schematics of the stochastic model and geometry of the cells and colonies within this model.	For the stochastic model, we assume that the shape of cells and colonies can both be approximated as spheres. For simplicity, we will talk about cells in the following. (a) Sketch of a cell on top of a substrate. It possesses multiple pili (gray lines). For the black colored pilus, the position is shown for three different time points. At $t_1$, the pilus has the length $\lmin$ and reaches the substrate. The pilus continues to slide along the substrate by modeling it as an infinitely stiff polymer with negligible pivotal stiffness. It stochastically binds to the substrate at time $t_3$. At this time it has the length $\length$. (b) Sketch of the quantities needed to determine the pili length $\Lproj$, projected on the surface, and the angle at which an attached pilus protrudes from the cell, $\aangle$. (c) Sketch of a cell in the stochastic model. In this example, the cell possesses three pili in total. The number of pili bound to the left and right sides are given by $\Nleft$ and $\Nright$. The pili can constantly change their states due to attachment, detachment, and the motion of the cell. (d) Allowed states and transitions in the model for a cell with a total number of three pili. The arrows show the transition between the different states ${(\Nleft,\Nright)}$. In the model, a cell is moving if a nonzero number of pili is attached only to one side. (e) Sketch of the transition of pili states due to motion over a pilus attachment point. Initially, all pili are on the right side and pull on the cell. Due to the motion of the cell and the different distances of the pilus attachment points, at some time point, the cell moves completely over one of the pili. This pilus then switches to the opposite side. 
		}
	\end{figure*}
	
	In the stochastic model, a cell is assumed to have a spherical shape with radius $\Radius$ sitting directly on a substrate (see Fig.~\ref{figstochastic}a). The simple spherical geometry enables us to estimate the parameters of the one-dimensional stochastic model from the binding dynamics of pili, originally a three-dimensional system. 
	A cell possesses $\Ntotal$ pili that protrude from a surface point characterized by a polar angle $\SurfAngle$ (see Fig.~\ref{figstochastic}a). 
	In Appendix~\ref{sec:appendixGeometricModel}, we develop a theoretical model of pili to analyze the properties of the pilus attachment points. In this model, the pili protrude from the surface of a spherical cell with the velocity $\vpro$ until they reach a length corresponding to an exponential length distribution with the mean length $\lchar$~\cite{holz:2010, zaburdaev:2014} and then start to retract with the velocity $\vret$. Both velocities have approximately the same value, thus we set ${\vpro = \vret}$~\cite{maier:2015}. Upon contact, the pilus slides along the surface. The tip of the pili can attach to the substrate with a rate $\Katt$. Attached pili protrude, on average, from the cell surface at an angle $\amean$ (see Fig.~\ref{figstochastic}b). The pili attach with an effective rate $\Keff$ to the substrate at a point that is defined by the mean $\Lmean$, projected on the substrate (see Fig.~\ref{figstochastic}b). 
	The effective attachment rate $\Keff$, the mean angle $\amean$, and the mean length $\Lmean$ of the pili result from the interplay of the geometry of a spherical cell sitting on top of a substrate and the growth dynamics of pili.
	We find that the probability of attachment for a single pilus is given by 
	\begin{equation}\label{eq:AttRatio}
	\Patt(R) = \frac{2 \lchar}{2 \lchar + \latt} \left[ 1 + \frac{\Radius}{\lchar} \exp \left( \frac{\Radius}{\lchar} \right) \mathrm{Ei} \left(-\frac{\Radius}{\lchar}\right)\right]
	\end{equation}  
	with the attachment length ${\latt = \vret \slash \Katt}$, denoting the mean distance a pilus protruding with velocity $\vret$ and attaching to a substrate with the rate $\Katt$.
	The larger a cell or the shorter the pili, the less likely it is for them to attach to a substrate. 
	From this probability and the time it takes for a pilus to reach the substrate, we estimate the effective attachment rate $\Keff$ of a single pilus and find (see Appendix~\ref{sec:appendixGeometricModel})
	\begin{equation}\label{eq:effattrate}
	\Keff(R) = \Gamma(R) \vret \frac{\latt - \Radius + \Radius \ln \left( \frac{\Radius}{\latt}\right)}{\left( \latt - \Radius\right)^2}.
	\end{equation}
	Similarly to the probability of attachment~$\Patt(R)$, the effective attachment rate $\Keff(R)$ of a single pilus of a cell is decreasing with increasing cell size or decreasing mean pili length. 
	A related calculation also provides us with the mean angle $\amean$ and the mean projected length $\Lmean$.
	
	\begin{table}[!tb]
		\caption{\label{table_parameters}%
			Given are the effective attachment rate~$\Keff$ of the pili and the mean projected length~$\Lmean$ and the mean angle~$\amean$ of attached pili. For the computation of these parameters we take parameters from Table~\ref{table_parametersinglecell}.
		}
		\begin{ruledtabular}
			\begin{tabular}{lcr}
				\textrm{ }&
				\textrm{fast}&
				\textrm{slow}\\
				\colrule
				$\Keff\ [\mathrm{Hz}]$  & 1.95 & 0.09  \\ 
				$\Lmean\ [\mu \mathrm{m}]$ & 0.59 & 1.58  \\ 
				$\amean\ $ & 2.03 & 1.78  \\ 
			\end{tabular}
		\end{ruledtabular}
	\end{table}
	
	The resulting parameter values for the motion of a single cell are shown in Table~\ref{table_parameters} for the two parameter sets from Table~\ref{table_parametersets}. Further parameters required for the calculation of the values given in Table~\ref{table_parameters}, specifically the cell radius~$\Radius$, the protrusion and retraction velocity~$\vpro$ and~$\vret$, and the mean pili length~$\lchar$ are given in Table~\ref{table_parametersinglecell}. 
	
	\begin{table}[b]
		\caption{\label{table_parametersinglecell}%
			Parameters of the stochastic model for the description of single cells.
		}
		\begin{ruledtabular}
			\begin{tabular}{lcr}
				\textrm{Name }&
				\textrm{Value}&
				\textrm{Ref.}\\
				\colrule
				Radius of single cell $\Radius$\ $[\mu \mathrm{m}]$  & 0.7 & \cite{zaburdaev:2014}  \\ 
				Number of pili $N$ & 15 & \cite{zaburdaev:2014} \\ 
				Translational mobility\ $\mobtrans$ $[\mu\mathrm{m} \slash (\mathrm{s\ pN})]$ & 10 & \cite{marathe:2014}  \\ 
				Protrusion velocity\ $\vpro$ $[\mu\mathrm{m} \slash \mathrm{s}]$ & 2 & \cite{maier:2013}  \\ 
				Retraction velocity\ $\vret$ $[\mu\mathrm{m} \slash \mathrm{s}]$ & 2 & \cite{maier:2013}  \\ 
				Mean pili length\ $\lchar$ $[\mu\mathrm{m}]$ & 1.5 & \cite{holz:2010, zaburdaev:2014}  \\ 
				Pili stalling force\ $\Fstall$ $[\mathrm{pN}]$ & 180 & \cite{maier:2013, marathe:2014}  \\ 
			\end{tabular}
		\end{ruledtabular}
	\end{table}
	
	Since an attached pilus emerges at an angle~$\amean$ from the surface of the cell, the force of a pilus has a normal component, pulling the cell towards the substrate, and a tangential component, mediating the motion of the cell. The normal force mediates sliding friction with the substrate, with the coefficient~$\slidcoeff$~\cite{persson:2000}. The sliding friction counteracts the tangential forces pulling on the cell and alters the force balance. To highlight this point, we consider a cell that possesses only one pilus which is attached and mediates a force $F$. The absolute value of the normal force component relative to the substrate is given by ${\Forcenorm} = - F\cos \amean$ and the absolute value of the tangential component is given by ${\Forcepara = F \sin \amean}$. Then, the total force that acts on the cell due to this particular pilus in the tangential direction relative to the substrate is given by ${\max \left[ 0,\ \Forcepara - \slidcoeff \Forcenorm \right]}$. In the model, we neglect torques and resulting rotations of the cell.
	
	We can now construct the master equation that describes the motility of cells on a substrate in one dimension.
	
	\subsection{\label{sec:OneDModel} Master equation for the stochastic model}
	
	In the stochastic model, a cell is modeled as a spherical particle moving through a viscous fluid with the translational mobility ${\mobtrans}$. It has multiple pili on its left and right sides that stochastically bind to and unbind from the substrate (see figure~\ref{figstochastic}c). 
	Attached pili generate pulling forces that act on the cell. We neglect the dynamics needed to reach the stationary state of the pilus pulling force, since the pilus spring constant~$\kspring$ is very high, allowing us to ignore its relaxation dynamics~\cite{biais:2010}.
	
	The cell has a probability~$\Plr(\Nleft,\Nright)$ to have $\Nleft$ pili attached to the left side and $\Nright$ pili attached to the right side. Due to the binding and unbinding dynamics, the cell switches between different states, thus affecting the probability of each state. 
	To characterize the dynamics of the probability~$\Plr(\Nleft,\Nright)$, we use a master equation approach. The master equation incorporates the following processes that can mediate transitions between states (see Fig.~\ref{figstochastic}d): attachment of free pili to the substrate, load-dependent detachment of attached pili and transitions due to the motility of cells over the pili attachment points.
	
	The dynamics of the probability $\Plr (\Nleft,\Nright)$ of the state ${(\Nleft,\Nright)}$, is described by the master equation  
	\begin{equation}\label{eq:mastereq}
	\frac{\mathrm{d}\Plr(\Nleft,\Nright)}{\mathrm{d}t} = \sum_{i=0}^{\Ntotal} \sum_{j=0}^{\Ntotal} \TransitionMatrix_{\Nleft \Nright}(i,j) \Plr(i,j),
	\end{equation}
	where the transition matrix $\TransitionMatrix$ is constructed from the rates corresponding to the transitions between the states ${(\Nleft,\Nright)}$.
	We will now define these rates and describe the emerging motion of cells.
	
	\subsubsection{\label{sec:Attachment} Attachment}
	A single pilus attaches to the substrate with an effective rate $\Keff$.  The transitions corresponding to the attachment of pili to the left side are given by ${(\Nleft,\Nright) \rightarrow (\Nleft+1,\Nright)}$ and ${(\Nleft,\Nright) \rightarrow (\Nleft,\Nright+1)}$ for binding to the right side.
	For an individual cell, the number of non-attached pili $\Nfree$ is given by
	\begin{equation}
	\Nfree = \Ntotal - \Nleft - \Nright, 
	\end{equation} 
	and we assume that whether a pilus will attach to the left or right side is not dependent on the current state ${(\Nleft,\ \Nright)}$.
	The transition rates for a single cell are then given by
	\begin{equation}
	\kattlcell = \kattrcell = \Nfree \Keff,
	\end{equation}
	where $\kattlcell$ and $\kattrcell$ are the rates for any pilus to attach to the left and right sides of the cell respectively. 
	
	\subsubsection{\label{sec:Detachment} Detachment}
	
	The detachment of a pilus is load-dependent (see Eq.~\eqref{eq:detrate}). Thus, we first need to calculate the forces acting on the pili on both sides (see Appendix~\ref{sec:piliforces}). 
	The characteristic force of a pilus is its stalling force $\Fstall$ (see Eq.~\eqref{eq:stalling}). A pilus retracts until it reaches this force and can only exceed this force if further pili are involved and pull against it. 
	If all attached pili in the system have a pulling force exceeding or equal the stalling force and all forces (specifically the tangential pili forces, the sliding friction forces with the substrate and  viscous friction forces with the fluid) balance, a stationary state is reached.
	
	We can then compute the stationary state of the pulling forces $\Fleftsteady$ and $\Frightsteady$ for a cell and find that for the case ${\Nleft=\Nright}$, the pulling forces on the left and right sides fulfill
	\begin{equation}
	\Fleftsteady (\Nleft = \Nright) = \Fstall
	\end{equation}
	and
	\begin{equation} 
	\Frightsteady (\Nleft = \Nright) = \Fstall.
	\end{equation}
	If pili are attached on both sides, but the case ${0 < \Nleft < \Nright}$ is fulfilled, the forces are given by
	\begin{eqnarray}
	\Fleftsteady (\Nleft,\Nright) &=& \max \left[ \Fstall,\ \Fstall \frac{\Nright}{\Nleft} \frac{1 + \slidcoeff \cot \amean}{1 - \slidcoeff \cot \amean} \right], \\
	\Frightsteady (\Nleft,\Nright) &=& \Fstall.
	\end{eqnarray}
	Due to symmetry, an equivalent equation can be given for states ${0 < \Nright < \Nleft}$. 
	
	If pili are only attached to the left side of the cell, the stationary force is given by
	\begin{equation}\label{eq:Forceleftoneside}
	\Fleftsteady(\Nleft,0) = \begin{cases}
	\frac{\Fstall}{1 + \frac{\Fstall \mobtrans \Nleft}{\vret} \left( 1 + \slidcoeff \cot \amean \right)}& \text{if } \left( 1 + \slidcoeff \cot \amean \right) \geq 0\\
	\Fstall              & \text{if } \left( 1 + \slidcoeff \cot \amean \right) < 0
	\end{cases}
	\end{equation}
	with the pilus retraction velocity $\vret$, the stalling relation (see Eq.~\eqref{eq:stalling}) and the mobility $\mobtrans$ of the cell moving through a viscous fluid. Here, the pilus pulling force is balanced by the sliding friction force and the hydrodynamics friction with the translational mobility~$\mobtrans$.
	The case where pili are only attached to the right side is equivalent.
	
	The transition rates for the transitions ${(\Nleft,\Nright) \rightarrow (\Nleft-1,\Nright)}$ and ${(\Nleft,\Nright) \rightarrow (\Nleft,\Nright-1)}$ are finally given by
	\begin{equation}
	\kdetl = \Nleft \Kdet \left( \Fleftsteady \right) 
	\end{equation}	
	and
	\begin{equation}
	\kdetr = \Nright \Kdet \left( \Frightsteady \right),
	\end{equation}	
	where the detachment rate $\Kdet$ is taken from Eq.~\eqref{eq:detrate}. 
	
	\subsubsection{\label{sec:MotionTransition} Transitions of states $(\Nleft,\Nright)$ due to cell motion}
	Cells can move on the substrate. The mean displacement that an individual pilus can mediate, here called $\Lmean$, is limited by the pilus length. As depicted in Fig.~\ref{figstochastic}e, a cell can, at some point, move over a pilus-substrate attachment point. If, for example, all pili are initially attached to the right side and the cell is moving over a pilus, then this corresponds to the transition ${(0,\Nright) \rightarrow (1,\Nright-1)}$. If we assume that the distance from the pilus-substrate attachment points to their anchor points (the connections to the cell surface) is exponentially distributed with the mean length $\Lmean$, then the corresponding transition rate~$\kmotr$ is given by
	\begin{equation}
	\kmotr = \Nright\frac{\vcell}{\Lmean}. 
	\end{equation}
	Here, $\vcell$ is the velocity of the cell. Its dependence on the state $(\Nleft,\Nright)$ is discussed in the following paragraph. The transition due to motion over pili on the left side, ${(\Nleft,0) \rightarrow (\Nleft-1,1)}$ with the transition rate~$\kmotl$, is treated analogously.	
	
	\subsubsection{\label{sec:Motionofcells} Motion of cells}
	
	Finally, it is necessary to connect the states ${(\Nleft, \Nright)}$ to the motion of the cells. Cells  move if pili are attached to one side, corresponding to states $(\Nleft>0,0)$ or $(0,\Nright>0)$ (see Fig.~\ref{figstochastic}d). 
	The cell velocity for these states is calculated in Appendix~\ref{sec:piliforces}.
	
	We found that if for example $\Nleft$ pili are attached on the left side, the cell moves in the left (and negative) direction with the velocity
	\begin{equation} \label{eq:vleft}
	\vcell(\Nleft,0) = -\begin{cases}
	\mobtrans \Nleft \Fleft \sin \amean & \text{if } \left( 1 + \slidcoeff \cot \amean \right) \geq 0\\
	0              & \text{if } \left( 1 + \slidcoeff \cot \amean \right) < 0
	\end{cases}
	\end{equation}
	where $\Fleft$ is given by Eq.~\eqref{eq:Forceleftoneside}. The term ${\sin \amean}$ results from the fact that only the force component tangential to the substrate can mediate the motion. If the sliding friction coefficient $\slidcoeff$ or the angle $\amean$ are high enough, the cell is no longer able to move.  The velocity has an equivalent form if pili are only attached to the right side but in this case the cell moves in the right direction and its velocity is positive. 
	
	\subsubsection{\label{sec:Colonymodification} Modifications of the stochastic model for colonies}
	
	To a large extent, colonies are treated similarly to individual cells within this model. The colonies are approximated as spheres with radius $\Radius$. While cells possess a fixed number $N$ of pili, the pili number is a function of the radius $\Radius$ for colonies and proportional to its surface area. This is realized by the surface density $\surfdens$ which gives the pili number from the proportionality to the surface area of the colony, ${N=4 \pi \surfdens \Radius^2}$. We choose the density $\surfdens = {0.8\ \mu \mathrm{m}^{-2}}$ that corresponds to $10$ pili for a cell of radius ${1\ \mu \mathrm{m}}$ and is in the order of magnitude of experimental values of single cells~\cite{holz:2010,zaburdaev:2014}.
	\begin{figure}[tb]
		\includegraphics[width=8cm]{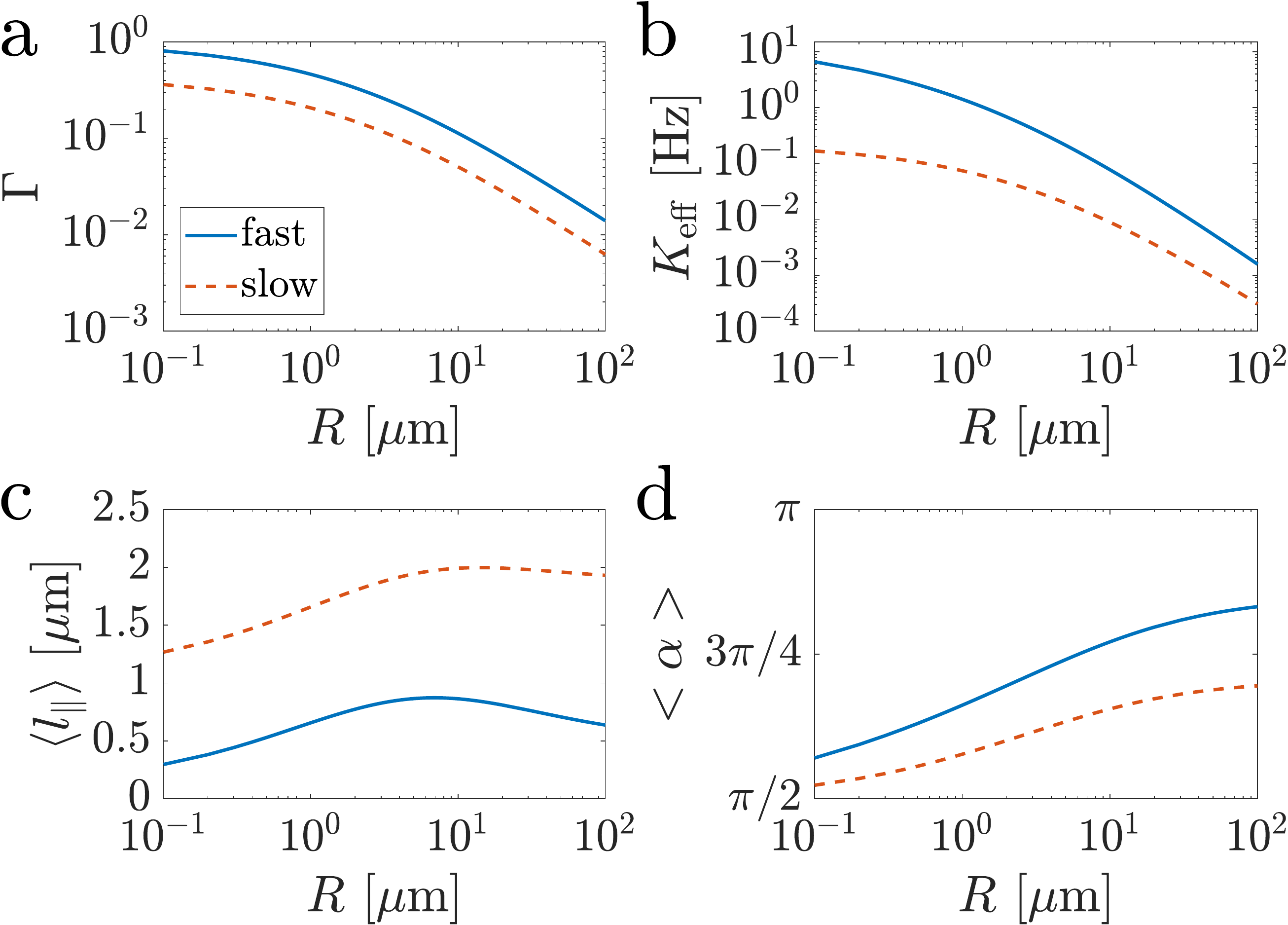}%
		\caption{\label{figgeometry} Colony size dependence of the effective attachment rate, the pili binding position and the mean attachment angle as a function of the colony radius~$\Radius$ for the \textit{fast} (solid blue line) and \textit{slow} (dashed red line) parameters. The values were calculated in Appendix~\ref{sec:appendixGeometricModel}. (a) Colony size dependence of the probability of pili attachment (see Eq.~\eqref{eq:AttRatio}) . (b) Size-dependence of the effective attachment rate, see Eq.~\eqref{eq:effattrate}. (c) Mean length of pili, projected on the substrate. (d) Mean angle of pili protruding from the cell surface.  For (a-d) we study the parameter sets given in Tables~\ref{table_parametersets} and~\ref{table_parametersinglecell}.}
	\end{figure}
	
	A major difference between colonies and cells can be found for the attachment dynamics.
	While we assume that pili are randomly distributed on the surface of a single cell, and, for example, can be all located on one side, this is not necessarily the case for a colony that is composed of several individual cells. All cells of the colony possess pili, independent of their position. For large colonies that possess a high number of pili, we can approximate the number of pili on each side (left and right) by ${\Ntotal \slash 2}$ and the attachment rates on both sides can differ because for the state ${(\Nleft,\ \Nright)}$ the colony has ${\Ntotal \slash 2 - \Nleft}$ free pili on the left side and ${\Ntotal \slash 2 - \Nright}$ on the right side. The corresponding attachment rates $\kattlcolony$ and $\kattrcolony$ for colonies are then given by   
	\begin{equation}
	\kattlcolony = \left( \frac{\Ntotal}{2} - \Nleft \right) \Keff 
	\end{equation}
	and
	\begin{equation}
	\kattrcolony = \left( \frac{\Ntotal}{2} - \Nright \right) \Keff. 
	\end{equation}
	The detachment dynamics and the motility of colonies is otherwise described in the same manner as for the case of an individual cell.
	Importantly, the probability of attachment $\Patt(R)$, the mean attachment rate $\Keff$, the projected length of an attached pilus $\Lmean$ and the attachment angle~$\amean$ are dependent on the radius $\Radius$ of the spherical colonies, as shown in Fig.~\ref{figgeometry}.
	\newline
	\newline
	Our stochastic model shares similarities to work previously published by M\"uller et al.~\cite{muller:2008} with respect to the attachment and detachment dynamics. They studied the bidirectional transport of cargo due to two populations of molecular motors, kinesin and dynein, pulling in different directions and having load dependent transport properties. Here, we additionally include the motility of the cell and colonies, its influence on the pili dynamics, the effects emerging from the sliding friction with the substrate and effects from the finite length of the pulling entities, the pili.
	
	We can numerically solve the master equation (see Eq.~\eqref{eq:mastereq}) by employing the  Gillespie algorithm~\cite{gillespie:1976,gillespie:1977} and thus model the emerging trajectory of a cell or a colony.
	
	\section{\label{sec:Results} Results}
	First, we want to demonstrate that a tug-of-war mechanism is sufficient to explain persistent motion of single cells with a characteristic length that exceeds the apparent pili length, as was previously shown experimentally~\cite{holz:2010, marathe:2014}. 
	Next, we study how the motion of individual cells is affected by the number of individual pili and find that, in agreement with experiments~\cite{holz:2010, marathe:2014}, the motility of cells is increased for higher number of pili. 
	Finally, we consider the motion of larger aggregates and emphasize the importance of sliding friction.
	
	\subsection{\label{sec:result1} Persistent motion of single cells}
	
\begin{figure*}[!tb]
		\includegraphics[width=11cm]{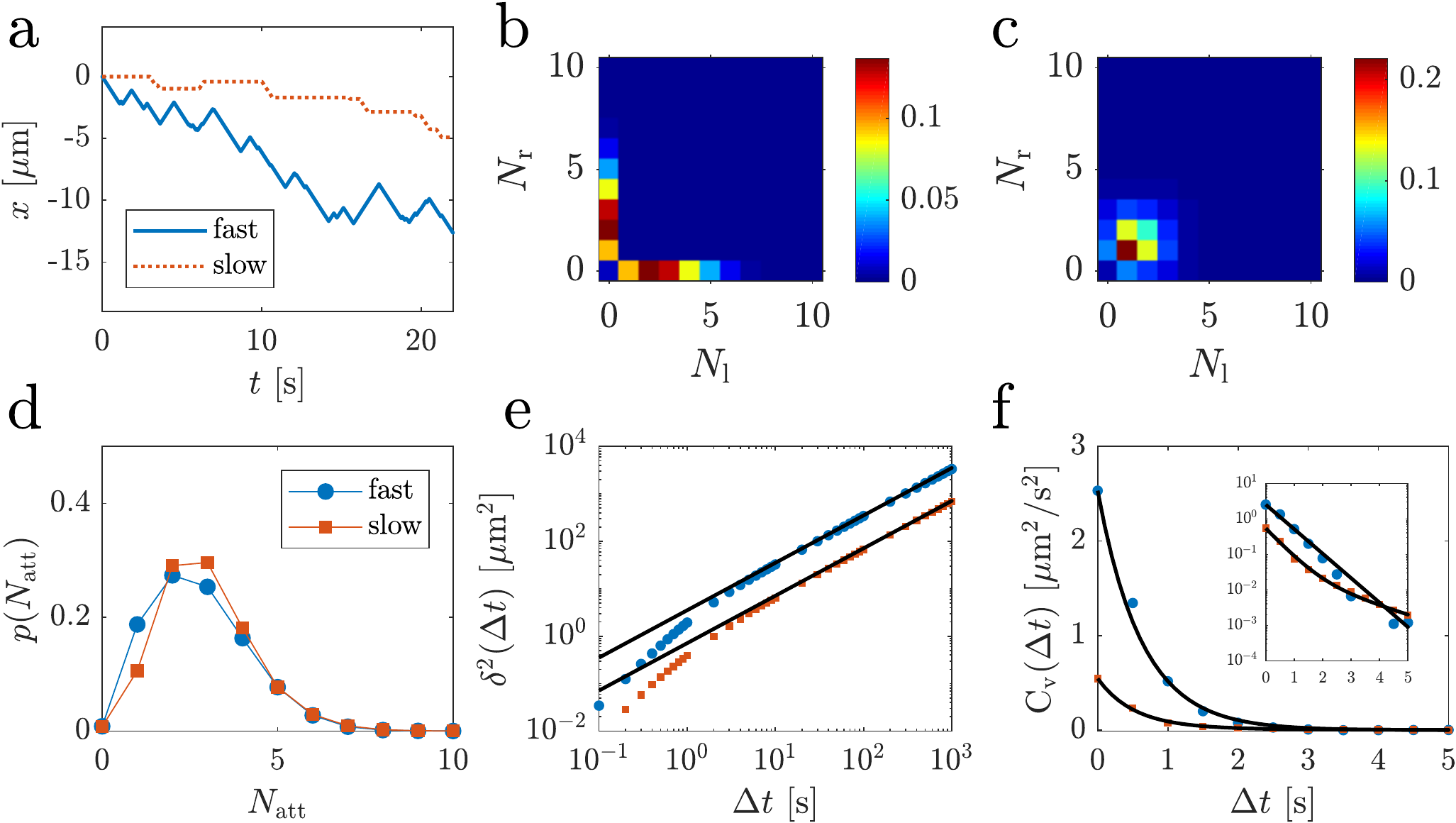}
		\caption{\label{figsinglestoch} Properties of the substrate motion of \textit{in silico} cells for the stochastic model. Data for trajectories of the \textit{fast} parameter set are shown in blue, data for the \textit{slow} parameter set are shown in red. (a) One-dimensional trajectories for the two parameter sets. (b) Probability of states of a single cell for the stochastic model, for the \textit{fast} parameter set. (c) Probability of states of a single cell for the stochastic model, for the \textit{slow} parameter set. (d) Histogram of the number of attached pili. (e) Time-averaged mean squared displacement. The black lines correspond to fits of Eq.~\eqref{eq:MSDstoch}. (f) Time-averaged velocity autocorrelation. The black lines correspond to fits of Eqs.~\eqref{eq:VACFstoch2} and~\eqref{eq:VACFstoch3}. The inset shows a logarithmic scaling for the velocity autocorrelation to highlight the exponential decay.}
	\end{figure*}

	For the analysis of single cell motion in the stochastic model, we simulated ${2 \times10^7}$ Gillespie steps of a single trajectory for the \textit{slow} and \textit{fast} parameter sets (see Table~\ref{table_parametersets}), corresponding to experiments of a duration of at least ${10^5\ \mathrm{s}}$. Parts of the trajectories are shown in Fig.~\ref{figsinglestoch}a. The cell possessed~${N=15}$ pili and for simplicity, we neglect sliding friction by setting ${\slidcoeff=0}$, because for the \textit{fast} and \textit{slow} parameter set, tangential forces clearly dominate over normal forces for small cell radii~$\Radius$. Further parameters of the stochastic model are given in Table~\ref{table_parameters}. 
	As motility of cells results from pili that are only attached at one side, we estimate which states ${(\Nleft,\Nright)}$ of attached pili had a higher probability (see Fig.~\ref{figsinglestoch}b and Fig.~\ref{figsinglestoch}c). 
	While mostly pili are only attached to one side of the cell for the \textit{fast} parameter set, for the \textit{slow} parameter set pili are often attached on both sides. Thus, we expect the \textit{fast} cells to be more motile. 
	From the probabilities of the different states ${(\Nleft, \Nright)}$, we can also estimate the probability distribution of attached pili (see Fig.~\ref{figsinglestoch}d). The distributions for both parameter sets shows a mean number of attached pili around 3-4, while on average more pili are attached for the \textit{slow} parameter set. In both cases, the number is considerably lower than the total number of pili~$N$.
	
Next to the statistics of the number of attached pili, we can also characterize the statistical properties of the trajectories of the cells. To this end, we calculated the time-averaged mean squared displacement (see Fig.~\ref{figsinglestoch}e)
\begin{equation}\label{eq:MSDstoch}
\mathrm{\delta^2}(\dt) = \langle \left[ \mathbf{r}(t+\dt) - \mathbf{r}(t)   \right]^2\rangle_t 
\end{equation}
for which the position vector is projected in one dimension, $x(t)$, and which fulfills $\mathrm{\delta^2}(\dt) \approx 2 D \dt$ for the long-time asymptotic. We found the diffusion coefficients ${D= (1.76 \pm 0.02 )\ \mu \mathrm{m}^2 \slash \mathrm{s}}$ for the \textit{fast} and ${D= (0.36 \pm 0.01 )\ \mu \mathrm{m}^2 \slash \mathrm{s}}$ for the \textit{slow} parameters. These results confirm that indeed cells for the \textit{fast} parameter set are more motile.
We can also compute the velocity autocorrelation function
	\begin{equation}\label{eq:VACFstoch}
	\mathrm{C_v}(\dt) = \langle \mathbf{v}(t+\dt) \mathbf{v}(t)   \rangle_t
	\end{equation}
	(see Fig.~\ref{figsinglestoch}f) where the velocities are one-dimensional vectors $v(t)$ and are defined by ${v(t) = [x(t) - x(t-\dt)] \slash \dt}$ with ${\dt=0.5\ \mathrm{s}}$.
		The same or higher $\dt$ were used in previous studies to characterize the velocity of \textit{N. gonorrhoeae}~\cite{holz:2010,zaburdaev:2014}. For the \textit{fast} parameter set we assume that the trajectory can be described by a persistent random walk,
		\begin{equation}\label{eq:VACFstoch2}
		\mathrm{C_v}(\dt) \simeq  \vc^2 \exp \left( - \frac{\dt}{\tc} \right),
		\end{equation}
		with the characteristic velocity~$\vc$ and the characteristic time~$\tc$. We found ${\tc =(0.62\pm0.08)\ \mathrm{s}}$ and ${\vc =(1.59\pm0.05) \ \mu \mathrm{m} \slash \mathrm{s}}$, allowing to compute the persistence length ${\lc = \vc \tc}$ that has the value ${\lc=0.99\ \mu \mathrm{m}}$.
		
	For the \textit{slow} parameter set, we find that a better fit for the velocity autocorrelation is given by a double exponential function, 
		\begin{equation}\label{eq:VACFstoch3}
		\mathrm{C_v}(\dt) \simeq  v_b^2 \exp \left( - \frac{\dt}{\tau_b} \right) + \vc^2 \exp \left( - \frac{\dt}{\tc} \right),   
		\end{equation}
		introducing a second characteristic velocity~$v_b$ and characteristic time~$\tau_b$. By fitting for this function, we find the characteristic times ${\tau_b =(0.51\pm0.09)\ \mathrm{s}}$ and ${\tc =(1.71\pm1.21)\ \mathrm{s}}$ and the characteristic velocities ${v_b =(0.71\pm0.07) \ \mu \mathrm{m} \slash \mathrm{s}}$ and ${\vc =(0.19\pm0.10) \ \mu \mathrm{m} \slash \mathrm{s}}$. The characteristic lengths are then given by ${l_b=v_b \tau_b=0.36\ \mu \mathrm{m}}$ and ${\lc=0.32\ \mu \mathrm{m}}$. The short time~$\tau_b$, but fast velocity~$v_b$ of the first exponential term originate from the motion of the cell in one direction that is interrupted by pili attaching to both sides and by motion over attached pili, but persistent. The second term, characterized by a higher characteristic time~$\tc$ and a smaller velocity~$\vc$, originates from changes of the direction of motion. The first term is not appearing for the \textit{fast} parameter set since states where pili are attached to both sides are appearing rarely and are, due to the high detachment rates, only short-lived (see Table~\ref{table_parametersets} and Fig.~\ref{figsinglestoch}). To further interpret the two exponential functions of the \textit{slow} parameter set, we compute their contribution to the diffusion coefficient from
		\begin{equation}\label{eq:MSDVACFLink}
		\mathrm{\delta^2}(\dt) = \int_{t_1=0}^{\dt}\mathrm{d}t_1\ \int_{t_2=0}^{\dt}\mathrm{d}t_2\ \mathrm{C_v}\left( \lvert t_1 - t_2\rvert \right)
		\end{equation}
		\cite{risken:1989} and given by ${D=D_b + D_c = v_b^2 \tau_b + \vc^2 \tc }$ for ${\dt \gg \tau_b}$ and ${\dt \gg \tc}$. We find ${D_b = 0.26\ \mu \mathrm{m}^2 \slash \mathrm{s}}$ and ${D_c = 0.06\ \mu \mathrm{m}^2 \slash \mathrm{s}}$, thus the first term (including $\tau_b$ and $v_b$) has the dominating contribution towards the diffusion coefficient.
	
		We can now compare the persistence lengths~$\lc$ and~$l_b$ to the apparent pili lengths that would be closest to the values measured experimentally and correspond to the projected pili length given in Table~\ref{table_parameters}. We find that the motion exhibits persistence over lengths higher than the mean projected pili length for the \textit{fast} parameter set (see Table~\ref{table_parameters}).  
	The stochastic model helps to see how the tug-of-war mechanism can mediate persistent motion. If, for example, more pili are located on the left side of a cell, then, due to the force balance, the pulling forces acting on the right pili are considerably higher. Then, the detachment rate of the pili located at the right side is also higher and they will detach with a higher probability than those on the left. Thus, states where pili are only attached to one side are preferred. 
	
\begin{figure*}[!tb]
		\includegraphics[width=12cm]{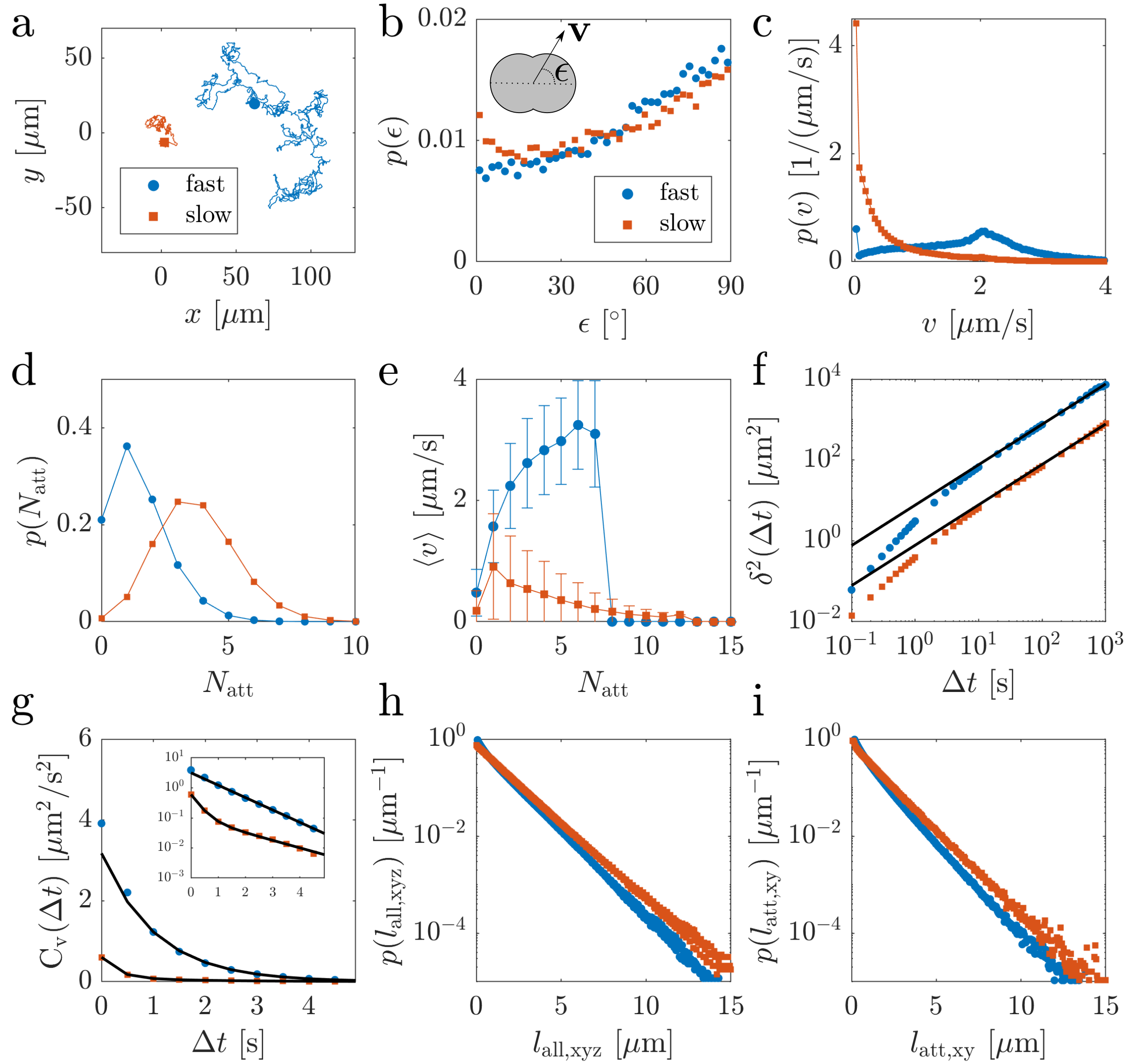}
		\caption{\label{figsinglecomp} Properties of the substrate motion of \textit{in silico} cells for the three-dimensional model. Data for trajectories of the \textit{fast} parameter set are shown in blue, data for the \textit{slow} parameter set are shown in red. Check Table~\ref{table_parametersets} and Appendix~\ref{sec:appendixCompModel} for the parameters. (a) Examples of trajectories projected on the substrate for a duration of ${1000\ \mathrm{s}}$. The dots represent the initial condition of the cells. (b) Distribution of angles~$\epsilon$ between the direction of motion projected on the substrate and the axis between the two cocci of the dumbbell-shaped cell. (c) Histogram of velocities. (d) Histogram of the number of attached pili. (e) Mean velocity of a cell as a function of the number of attached pili. (f) Time-averaged mean squared displacement. The black lines correspond to fits of the equation $\mathrm{\delta^2}(\dt) \approx 4 D \dt$. (g) Time-averaged velocity autocorrelation. The black lines correspond to fits of the Eqs.~\eqref{eq:VACFstoch2} and~\eqref{eq:VACFstoch3}. The inset shows a logarithmic scale of the velocity autocorrelation function. (h) Histogram of the contour length of all pili of a cell. (i) Histogram of the contour length projected on the substrate of attached pili. }
		
	\end{figure*}

	While the tug-of-war mechanism is strong enough for the \textit{fast} parameter set to generate a characteristic length of the motility that exceeds the projected pili length of ${\Lmean=0.59\ \mu \mathrm{m}}$, the length~$\lc$ does not exceed the characteristic pili length~${\lchar=1.5\ \mu \mathrm{m}}$ chosen as input in the system (see Table~\ref{table_parametersinglecell}). One reason for the discrepancy to the experiment is the one-dimensionality of the system which reduces the characteristic velocity and thus also the characteristic length of the movement. For example, the maximum velocity of a cell with two perpendicularly oriented and attached pili is $\sqrt{2}$ times greater than the velocity of a cell moving in one dimension due to the Pythagorean theorem. On the other side, in two dimensions the persistence might be reduced due to rotational diffusion of the cell.
	To see how these factors can affect the predictions of the stochastic model, we simulated the motility of individual cells with an advanced three-dimensional model (see~\cite{poenisch:2017} and Appendix~\ref{sec:appendixCompModel} for a description). This model allows for a microscopic description of the single cells and colonies interacting via pili with the substrate. It includes the explicit details of the cell geometry, the pili dynamics and the forces originating from attached pili and excluded volume effects with to the substrate. We sampled the trajectories of $50$ cells, each having a duration of ${1000\ \mathrm{s}}$. Examples for the trajectories of the two parameter sets \textit{fast} and \textit{slow} (see Table~\ref{table_parametersets}), projected on the substrate, are shown in Fig.~\ref{figsinglecomp}a. We first check if the three-dimensional model is capable to reproduce previously reported properties of single cell motion~\cite{holz:2010, zaburdaev:2014, marathe:2014}.
	For \textit{N. gonorrhoeae} cells it was shown before that cells preferentially move in the direction perpendicular to the long axis of the dumbbell-shaped cell~\cite{zaburdaev:2014}. When we determine the distribution of the angle $\epsilon$ between the long axis of the cell, the axis between the two cocci of the dumbbell-shaped cell, and the direction of motion projected on the substrate, we observe a similar behavior for both parameter sets (see Fig.~\ref{figsinglecomp}b). 
	Experimentally, it was reported that the cell velocity distribution exhibits a bimodal shape~\cite{zaburdaev:2014}. Indeed, for the \textit{fast} parameters we see qualitative agreement to this observation (see Fig.~\ref{figsinglecomp}c): the velocities do not only possess a peak at around ${2\ \mu \mathrm{m} \slash \mathrm{s}}$, corresponding to the retraction velocity $\vret$ of the pili, but also a peak at zero velocity, corresponding to no moving cells. Such a behavior is less pronounced for the \textit{slow} parameter set. Two different processes could lead to a vanishing velocity of a cell: either no pilus is attached to the substrate or the cell is trapped between several pili pulling against each other. From the distribution of attached pili (see Fig. ~\ref{figsinglecomp}d) we found that ${\sim20\ \%}$ of the time no pili are attached to the substrate for the \textit{fast} parameter set, explaining the peak at ${v=0}$ in Fig.~\ref{figsinglecomp}c since the cell is incapable of motion if no pili are attached. This finding agrees with Total internal reflection fluorescence (TIRF) microscopy data of \textit{N. gonorrhoeae} showing that for cells moving on a poly-D-lysine-coated surface in ${15 - 20\ \%}$ of the time no pili were observed in the vicinity of the substrate~\cite{eriksson:2015}.  For the \textit{slow} parameter set, pili are always attached to the substrate and we observe a distinct peak at ${v=0}$. 
		In contrast to the stochastic model, the three-dimensional model predicts that more pili are attached to the substrate for the \textit{slow} parameter set than for the \textit{fast} parameter set.
	Such behavior can result from the coarse-graining of the stochastic model. There, all pili belonging to one side of a cell are experiencing the same force, while in the three-dimensional model the forces are not necessarily equal. For example, in the three-dimensional model it is possible that only one pilus exerts the pulling force, even though other pili are attached on the same side. These pili will unbind with a lower probability than in the stochastic model and the number of attached pili can differ.
	We also find that on average only 0 to 5 pili for the \textit{fast} parameter set and 1 to 8 pili for the \textit{slow} parameter set are attached to the substrate, even though the cell possesses 15 pili. A similar behavior is observed in the stochastic model (see Fig.~\ref{figsinglestoch}d). In experiments typically $10-20$ are found for cells of \textit{N. gonorrhoeae}~\cite{zaburdaev:2014, marathe:2014, holz:2010}. With methods such as TIRF (due to the fact that imaging only takes place in vicinity of the substrate) and electron microscopy (due to the sample preparation) most likely one only images attached pili, thus we find that potentially the total number of attached and free pili of a cell is underestimated. 
	
	In the stochastic model, the total number of attached pili on one side of the cell affects the velocity of cells, as described by Eq.~\eqref{eq:vleft}. For the three-dimensional model, we observe the same qualitative behavior by estimating the average velocity of cells as a function of the number of attached pili for the \textit{fast} parameter set (see Fig.~\ref{figsinglecomp}e). The opposite behavior is observed for the \textit{slow} parameter set. This is in agreement with the stochastic model: pili are mostly attached on one side of the cell for the \textit{fast} parameter set and, thus, the velocity increases with a higher number of pili. For the \textit{slow} parameter set, pili are mostly attached equally on both sides, thus Eq.~\eqref{eq:vleft} does not apply.
	
We can also calculate the statistical properties of the trajectories from the three-dimensional model. From the two-dimensional mean squared displacement (see Eq.~\eqref{eq:MSDstoch} and Fig.~\ref{figsinglecomp}f) 
with its long-time asymptotic ${\mathrm{\delta^2}(\dt) \approx 4 D \dt}$ we then estimated the diffusion coefficient $D$ and found ${D= (1.91 \pm 0.03 )\ \mu \mathrm{m}^2 \slash \mathrm{s}}$ for the \textit{fast} parameter set and ${D= (0.19 \pm 0.02 )\ \mu \mathrm{m}^2 \slash \mathrm{s}}$ for the \textit{slow} parameter set. 
For small lag times $\dt$ we found that the mean-squared displacement exhibits super-diffusive behavior. We can interpret this behavior as a persistent random walk, as previously done for the motion of {\textit{N. gonorrhoeae}} bacteria~\cite{holz:2010, marathe:2014, bisht:2017}. 
The velocity autocorrelation in two dimensions is given by Eq.~\eqref{eq:VACFstoch}
	(see Fig.~\ref{figsinglecomp}g). Here, the velocity results from ${\textbf{v}(t) = \left[\textbf{r}(t) - \textbf{r}(t-\dt)\right] \slash \dt}$ with ${\dt=0.5\ \mathrm{s}}$. 
	For the \textit{fast} parameter set we describe the trajectory as a persistent random walk with an exponential behavior of the velocity autocorrelation (see Eq.~\eqref{eq:VACFstoch2})
	with the characteristic time $\tc$ and the characteristic velocity $\vc$. We find ${\tc =(1.05\pm0.03)\ \mathrm{s}}$ and ${\vc =(1.78\pm0.06) \ \mu \mathrm{m} \slash \mathrm{s}}$
	The characteristic length scale of the motion results from ${\lc = \vc \tc}$ and is given by ${\lc=1.87\ \mu \mathrm{m}}$.
	Again, the velocity autocorrelation function for the \textit{slow} parameter set can be described by a double exponential function, shown in Eq.~\eqref{eq:VACFstoch3}, with the characteristic times~$\tau_b$ and~$\tc$ and characteristic velocities~$v_b$ and~$\vc$. By fitting, we find ${\tau_b =(0.30\pm0.02)\ \mathrm{s}}$ and ${\tc=(1.73\pm0.27)\ \mathrm{s}}$ and ${v_b =(0.71\pm0.05) \ \mu \mathrm{m} \slash \mathrm{s}}$ and ${\vc =(0.32\pm0.14) \ \mu \mathrm{m} \slash \mathrm{s}}$. This corresponds to the characteristic length scales ${l_b= v_b \tau_b=\vc\tc = 0.21\ \mu\mathrm{m}}$ and ${\lc = 0.56\ \mu \mathrm{m}}$. Additionally, using Eq.\eqref{eq:MSDVACFLink}, we find that both exponential functions contribute equally to the diffusion coefficient for ${\dt \gg \tau_b}$ and ${\dt \gg \tc}$.
	
	As the mean pili length was chosen to be ${1.5\ \mu \mathrm{m}}$ and is smaller than the persistence length of the motility for the \textit{fast} parameters, this result is in agreement with the stochastic model. We also tested whether the persistence results from the diplococcus shape of the bacteria by computing the velocity autocorrelation for spherical cells. We found that the persistence becomes even larger with a characteristic length ${\lc=2.4\ \mathrm{\mu m}}$ for the \textit{fast} parameter set. This results most likely from the fact that a diplococcus can rotate such that its long axis is perpendicular to the substrate, and thus can be approximated by a sphere with a higher radius, but the same pili number. In this case, the effective attachment rate is lower, thus decreasing the motility of the cell (see Fig.~\ref{figgeometry}b).
	
	In the simulation, we set the characteristic length~$\lchar$ of free pili. Due to attachment and the triggered retraction, the experimentally measured pili lengths~\cite{zaburdaev:2014, marathe:2014} might differ from~$\lchar$. To investigate the pili length distribution, we tested different definitions of the pili length. Firstly, we just took the contour length of all pili $\lallxyz$, both attached and free. Alternatively, we tried to imitate the experiment, where the pilus length is measured by transmission electron microscopy~\cite{holz:2010, zaburdaev:2014}. We expect that the preparation of the samples affects the pili, especially those that were not attached to the substrate. They may break, bend or just detach from the cell. Thus, we were only taking those pili that were attached to the substrate. Additionally, in the experiment one is only measuring the pilus length projected on the substrate. Thus, we define the second length $\lattxy$ as the distance of the start and end point of attached pili projected on the substrate. The distribution of these lengths is shown in Figs.~\ref{figsinglecomp}h and~\ref{figsinglecomp}i. For the \textit{fast} parameters we observe mean lengths ${\langle \lallxyz \rangle = 1.13\ \mu \mathrm{m} }$ and ${\langle \lattxy \rangle = 0.91\ \mu \mathrm{m} }$. For the \textit{slow} parameters we find ${\langle \lallxyz \rangle = 1.35\ \mu \mathrm{m} }$ and ${\langle \lattxy \rangle = 1.16\ \mu \mathrm{m} }$. Thus, the observed persistent motion in the three-dimensional model is even more pronounced for the~\textit{fast} parameter set, since the measured pili lengths are considerably smaller than the characteristic length~${\lc=1.5\ \mathrm{\mu m}}$ of the motility.
	
	Taken together, we observe that the properties of cell motion in the three-dimensional model agree qualitatively and quantitatively with the key experimental observations. The stochastic model shows that a tug-of-war is the mechanism underlying the persistence of cell motion without invoking directional memory~\cite{marathe:2014}.
	
	\subsection{\label{sec:result2} Increasing the pili number enhances the substrate motility of single cells}
	
	Experimental studies with mutants of \textit{N. gonorrhoeae} have shown that the more pili a cell possesses, the more motile it gets~\cite{holz:2010}: with higher pili numbers, the persistence time $\tc$ is increasing. 
	We can now test if this observation can also be reproduced within the stochastic model by varying the total number of pili $N$. 
	For simplicity, we first neglect sliding friction by setting the sliding friction coefficient ${\slidcoeff=0}$ since we still consider single cells only and the tangential forces of the pili dominate over their normal force components. Since we found that only the \textit{fast} parameter set reproduces the experimentally observed features of single cell motility, in the following we only focus on this parameter set.
	
\begin{figure}[tb]
		\centering
		\includegraphics[width=8cm]{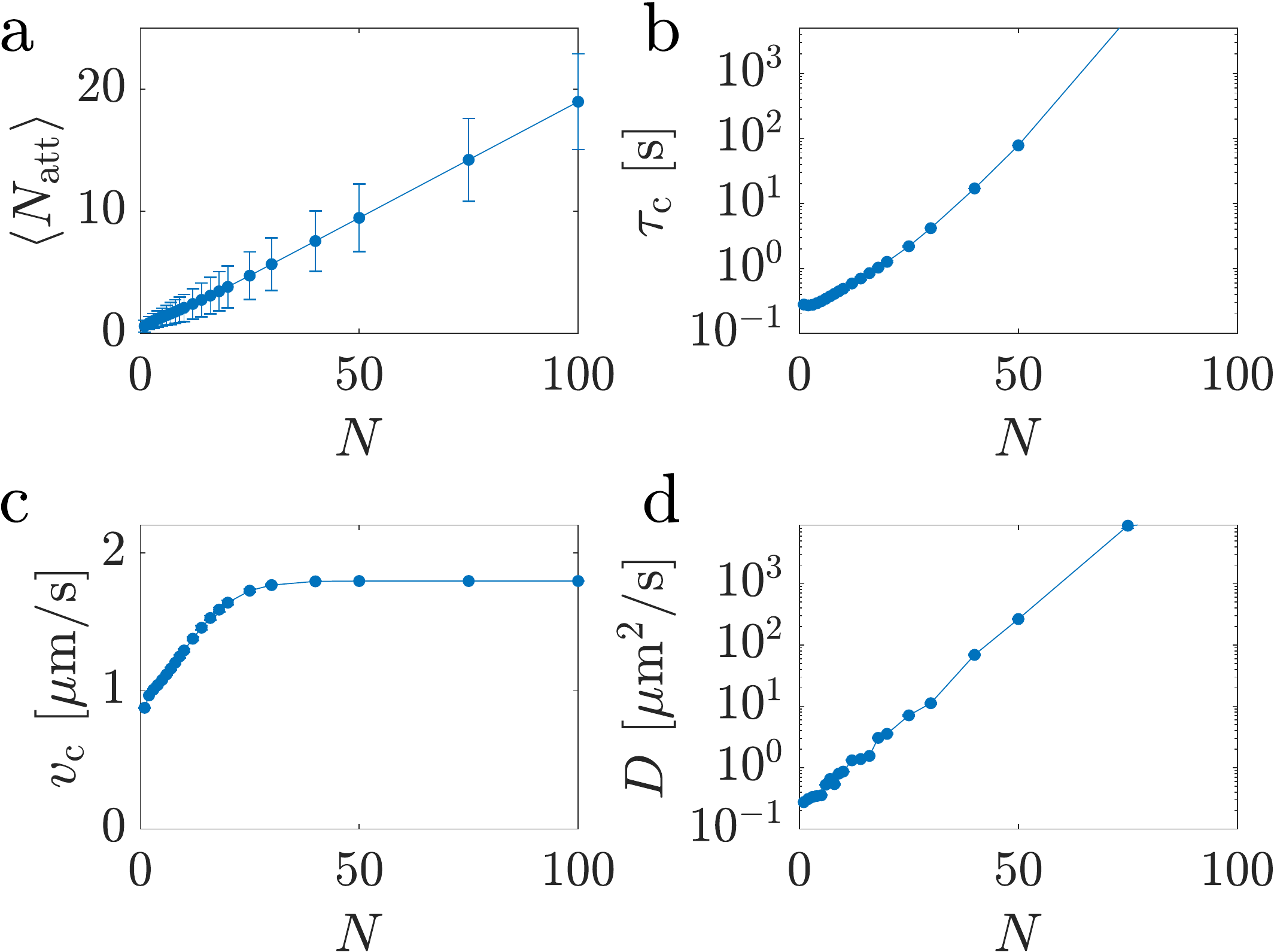}
		\caption{\label{fignumberstoch} Pili-number dependence of an \textit{in silico} cell as calculated from the stochastic model for the \textit{fast} parameter set. 
			(a) Mean number of attached pili as a function of the total number of pili. The error bars represent the variance. (b) Characteristic time $\tc$ as a function of the total pili number. (c) Characteristic velocity $\vc$ as a function of the total pili number. (d) Diffusion coefficient $D$ as a function of the pili number.}
	\end{figure}

	As shown in Fig.~\ref{fignumberstoch}, we observe qualitative agreement to the experiment. 
	The characteristic time, the characteristic velocity, the diffusion coefficient, and the mean number of attached pili are all increasing with increasing pili number. While the diffusion coefficient~$D$ and the characteristic time~$\tc$ increase exponentially with increasing pili number, the velocity~$\vc$ saturates due to the fact that retracting pili cannot exceed the characteristic pili velocity $\vret$. 
	Since the characteristic velocity~$\vc$ and the characteristic time~$\tc$ are both increasing with increasing pili number, the characteristic length~$\lc$ of the motion will also increase. In the previous section we showed that the average number of attached pili~$\Natt$ is lower than the total number of pili~$N$ (see Figs.~\ref{figsinglestoch} and~\ref{figsinglecomp}). In experiments, typically only attached pili are imaged, thus the real number of pili a cell possesses can be higher. This can explain why we find a characteristic length of the motion~$\lc$ that is smaller than the characteristic pili length~$\lchar$ for a cell possessing only ${N=15}$ pili.
	
\begin{figure}[tb]
		\centering
		\includegraphics[width=8cm]{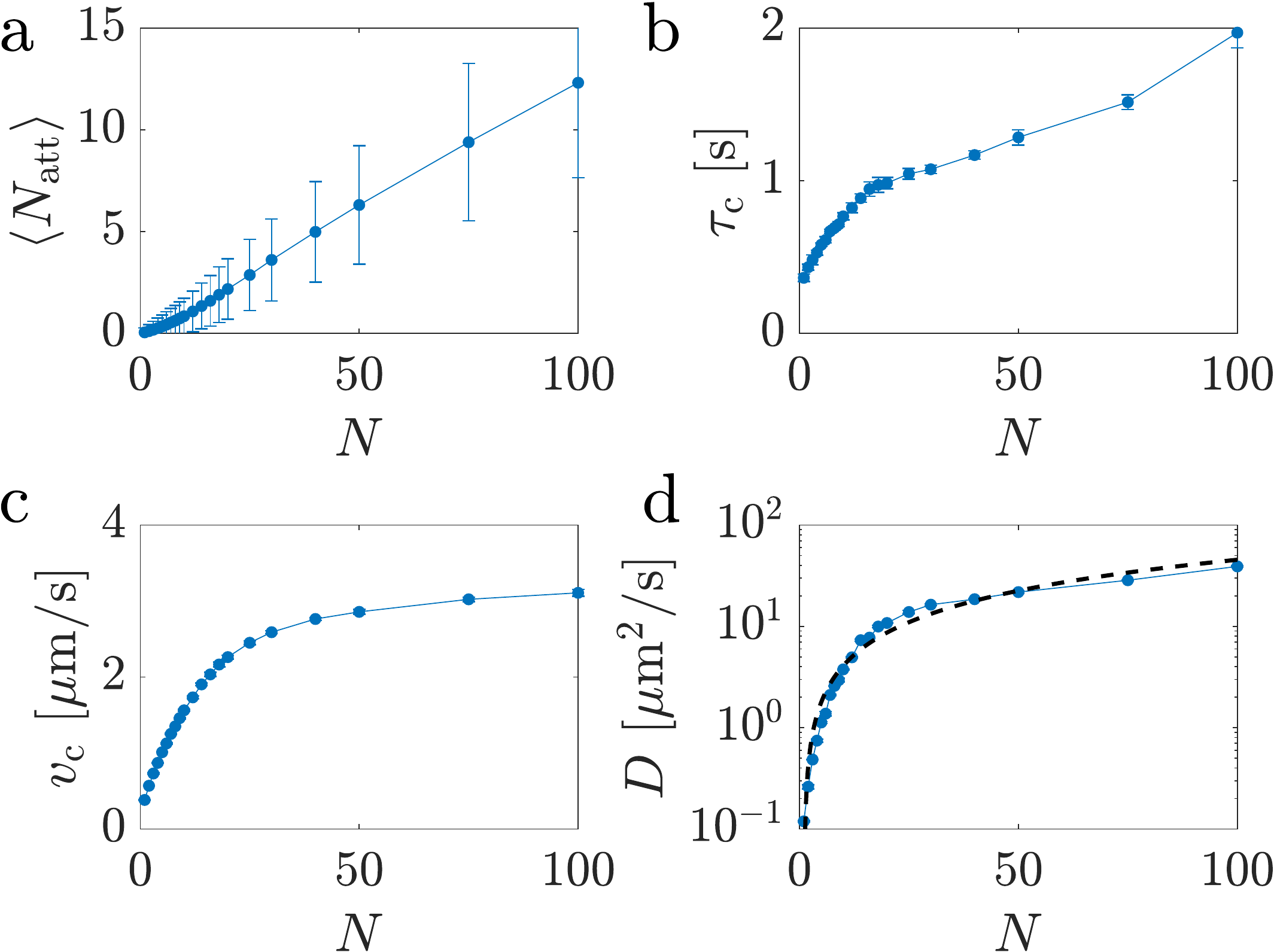}
		\caption{\label{fignumbercomp} Pili-number dependence of an \textit{in silico} cell as calculated from the three-dimensional model for the \textit{fast} parameter set. (a) Mean number of attached pili as a function of the total number of pili. (b) Characteristic time $\tc$ as a function of the total pili number. (c) Characteristic velocity $\vc$ as a function of the total pili number. (d) Diffusion coefficient $D$ as a function of the pili number. The black dashed line represents a scaling of the form ${D \propto N}$, as predicted in Appendix~\ref{sec:Map1Dto2D} and resulting from the rotational diffusion of the cell.}
	\end{figure}

	When we calculate the characteristic length scale ${\lc=\tc \vc}$ from the results of the stochastic model, we find that it increases rapidly over multiple orders of magnitude with increasing number of pili. The characteristic time $\tc$ and the diffusion coefficient $D$ show the same behavior and exceed the results from experiments~\cite{holz:2010,marathe:2014} considerably for high numbers of pili $\Ntotal>25$. We suggest that the exponential increase of the characteristic time~$\tc$ is originating from the trapping of a cell in a state where pili are only located on one side of the cell. To test this hypothesis, we treat the pili dynamics as a random walk with the number of attached pili $N$ as the location of the random walk and calculate the mean first passage time to the point ${N=0}$ corresponding to losing all attached pili. In Appendix~\ref{sec:MFPT} we show that this time indeed is increasing exponentially with pili number.
	One process that counter-acts the exponential increase of the characteristic time is the rotational diffusion of the cell, which would confine the diffusion of every persistent cells. 
	To test this hypothesis, we used our three-dimensional model and sampled the trajectories of $25$ cells for different numbers of pili $\Ntotal$. As shown in Fig.~\ref{fignumbercomp}, for the \textit{fast} parameter set the persistence time $\tc$, the mean number of attached pili, the characteristic velocity $\vc$ and the diffusion coefficient $D$ increase with increasing numbers of pili $\Ntotal$. This result is in qualitative agreement with the experiment and the results of the stochastic model. 
	Importantly, the values of $\tc$ and $D$ indeed do not reach as high values as those observed in the stochastic model. This behavior originates from the rotational diffusion of the cell. In Appendix~\ref{sec:Map1Dto2D}, we show how the scaling of the diffusion coefficient can be altered from an exponential scaling ${D \propto \exp(N \slash N_0)}$ (with the characteristic pili number~$N_0$) for a motion without rotational diffusion to ${D \propto N}$ by considering rotations of the direction of motion (see Fig.~\ref{fignumbercomp}d).
	The difference in the mean number of attached pili~$\langle \Natt \rangle$ of both models can be explained by the fact that we assume that for the stochastic model, all pili are on the lower hemisphere of the cell and thus, correspond to a cell with the doubled number of pili in the three-dimensional model, ignoring effects due to the rotation of the cell.
	Thus, also the pili number dependence of single cell motility can be recapitulated by the simple stochastic model.
	
	Finally, we checked whether the single cell motion is affected by sliding friction. We can expect that the sliding friction would only play a minor role due to the weaker normal component of the pili forces (see Fig.~\ref{figgeometry}d). Indeed, for the motility of single cell as a function of the number of pili (see Figure~\ref{fig:SS1} for the stochastic model and Figure~\ref{fig:SS2} for the three-dimensional model) we observe qualitative agreement to the behavior we previously showed for a sliding friction ${\slidcoeff = 0}$, shown in Fig.~\ref{fignumberstoch} and Fig.~\ref{fignumbercomp}. The motility of cells is indeed increasing with increasing pili number, even if there is friction between the individual cell and the substrate.
	
	In the next section we will apply the stochastic model to study how microcolonies, aggregates consisting of multiple cells, move on a substrate. 
	
	\subsection{\label{sec:result3} Colony size-dependent motility}
	
	In the following, for simplicity of the presentation we will only focus on the \textit{fast} parameter set. 
	\begin{figure}[tb]
		\includegraphics[width=8cm]{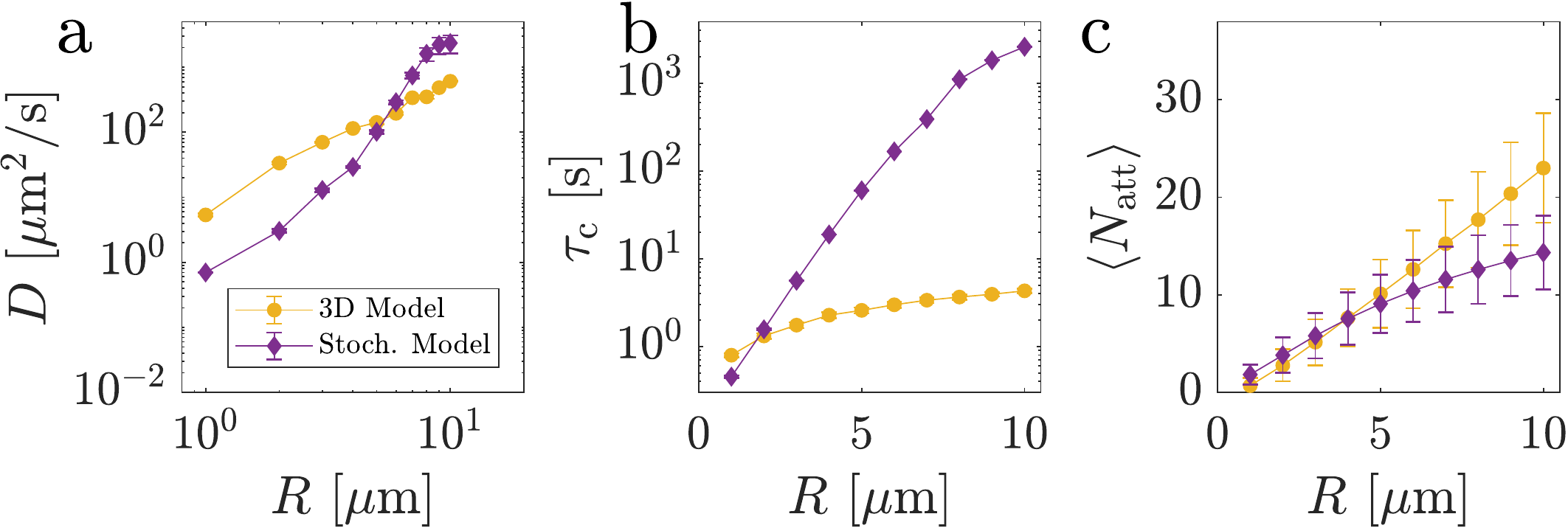}
		\caption{\label{figdiffnofric} Colony size-dependent motility of bacterial aggregates for the computational and stochastic models. (a) Diffusion coefficient of colonies described by the computational and stochastic model. For both, $D$ is increasing with increasing colony radius $\Radius$. (b) Characteristic time scale $\tc$ of the motion. In both models, it is increasing for larger colonies. (c) Mean number of attached pili as a function of the colony sizes. Larger colonies possess more pili. Thus, not surprisingly larger colonies also have more pili attached to the substrate. For (a-c) we used parameters given in Appendix~\ref{sec:appendixCompModel} and the \textit{fast} parameter set of Table~\ref{table_parametersets}. For the simulations of our stochastic model, we used the colony size dependence of the attachment rate and the mean displacement a pilus mediates as given in Figs.~\ref{figgeometry}c-f. }
	\end{figure}
	First, we simulated trajectories of colonies of radii between $1$ to ${10\ \mu \mathrm{m}}$ resulting from the stochastic model by considering the colony size-dependent behavior of the parameters, as shown in Fig.~\ref{figgeometry}b-d. If we ignore the sliding friction by setting ${\slidcoeff=0}$, we find that the diffusion coefficient (see Fig.~\ref{figdiffnofric}a) and the persistence of the motion increase with increasing colony sizes (see Fig.~\ref{figdiffnofric}b), contrary to experimental observations~\cite{taktikos:2015}, but in agreement with our results for the  motility of a single cell for higher numbers of pili (see Fig.~\ref{figdiffnofric}). 
	Indeed, if a tug-of-war mechanism drives the motility, the motility should be enhanced due to the increasing total number of pili of the colony (in agreement with the geometric model, see  Appendix~\ref{sec:appendixGeometricModel}).  
	
	To check this prediction, we used the three-dimensional model to simulate trajectories of microcolonies of different radii $\Radius$.  Each colony possesses ${4 \pi \surfdens \Radius^2}$ pili, with the pilus surface density $\surfdens = {0.8\ \mu \mathrm{m}^{-2}}$. For each colony size $\Radius$, we simulated $12$ trajectories, each of a total duration of $1000\ \mathrm{s}$. Importantly, also here we neglect the friction with the substrate.
	We calculated the diffusion coefficient $D$ (see Fig.~\ref{figdiffnofric}a), the characteristic time of motion $\tc$ (see Fig.~\ref{figdiffnofric}b) and the mean number of attached pili (see Fig.~\ref{figdiffnofric}c), which, in agreement with the predictions of the stochastic model, increase for larger colonies.  
	Thus, we see that a tug-of-war-mechanism together with the geometrical considerations for the colony parameters is not able to explain the experimentally observed decreasing motility of larger colonies.
	
\begin{figure}[tb]
		\includegraphics[width=8cm]{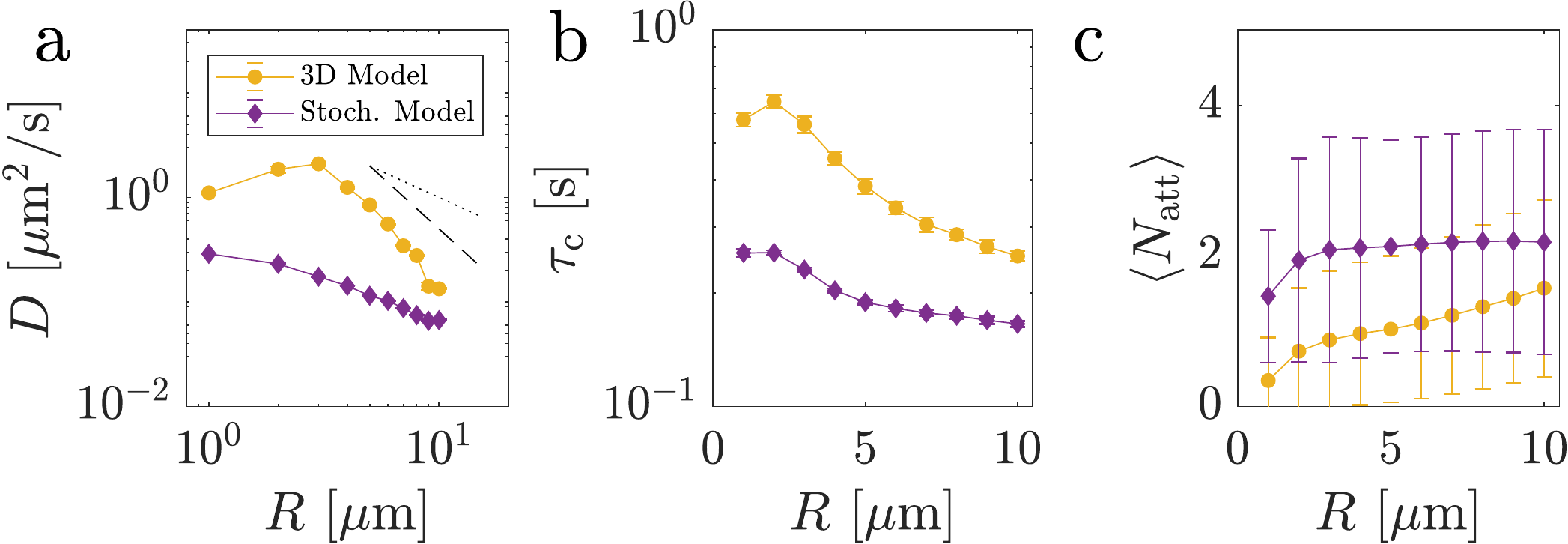}
		\caption{\label{figdifffric} Colony size-dependent motility of bacterial aggregates with substrate friction. (a) Diffusion coefficient of colonies described by the computational and stochastic models with sliding friction coefficients ${\slidcoeff = 0.8}$. The dotted line corresponds to a scaling ${D \propto R^{-1}}$, the dashed line corresponds to ${D \propto R^{-2}}$. (b) Characteristic time scale $\tc$ of the motion. (c) For ${\slidcoeff = 0.8}$ the number of attached pili is increasing. For (a-c) we used parameters given in Appendix~\ref{sec:appendixCompModel} and the \textit{fast} parameter set of Table~\ref{table_parametersets}. }
	\end{figure}

	We wondered which additional mechanism could explain the weaker motility of large colonies? So far, we only considered the hydrodynamic friction of microcolonies but ignored friction between the cells and the substrate completely. Now, we suggest that, in addition, one should consider such sliding friction~\cite{persson:2000}. We expect that as colony size grows, the pili-mediated normal force pulling the colony into the substrate increases and expect that the substrate friction would play an increasing role. In Fig.~\ref{figgeometry}d we show that the average angle between the pili and the substrate is increasing with increasing colony size. Thus, for larger colonies, the normal component of the pili forces will grow, while the tangential component that effectively causes the motion of the aggregates, will decrease. The higher the normal component of the force, the higher the sliding friction force that opposes the motion and the harder it is for the colony to move. To test this hypothesis, we studied the motion of microcolonies of different sizes for the sliding friction coefficient ${\slidcoeff = 0.8}$.
	First, we considered the role of sliding friction in the stochastic model. The sliding friction affects the forces with which pili are pulling on the colony in the stationary state. Colonies can only move if the condition ${1 + \slidcoeff \cot \amean > 0}$ is fulfilled (see Eq.~\eqref{eq:vleft}). Thus, we can calculate that for the sliding friction coefficient ${\slidcoeff = 0.8}$ and the \textit{fast} parameter set, colonies can only move if they have a size ${\Radius < 14.2\ \mu \mathrm{m}}$ (see Figure~\ref{fig:SS3}). 
	As shown in Fig.~\ref{figdifffric}a the diffusion coefficient is indeed decreasing for increasing colony sizes $\Radius$.
	
	From the velocity autocorrelation, we again calculated the characteristic time of the motion $\tc$ which is decreasing for increasing colony sizes (see Fig.~\ref{figdifffric}b), indicating that the colony motion becomes less persistent. At the same time, the mean number of attached pili is increasing (see Fig.~\ref{figdifffric}c). The average number for sliding friction is surprisingly small for colonies possessing up to thousands of pili. This discrepancy can be explained by considering that for the \textit{fast} parameter set pili bind frequently, but only weakly. While at the same time only a small number of pili are attached to the substrate, a larger amount is near the substrate, competing in the tug-of-war.
	Consistently, if we include sliding friction in the three-dimensional model, we find that the diffusion coefficient is decreasing for increasing colony size (see Fig.~\ref{figdifffric}a). Importantly, we see that by fitting a power law of the form ${R^{-\eta}}$ to $D(R)$ we find ${\eta > 1}$, in agreement with experimental observations~\cite{taktikos:2015}. The characteristic time $\tc$ and the mean number of attached pili $\langle \Natt \rangle$ also show a behavior that agrees with the stochastic model.
	
	To summarize, our stochastic and three-dimensional model show that for a tug-of-war mechanism alone, colonies exhibit an enhanced motility for increasing colony sizes. By introducing sliding friction, we can reproduce the experimentally observed decreasing motility with increasing colony size.
	
	\section{\label{sec:Discussion} Discussion}
	\begin{figure*}[!t]
		\includegraphics[width=12cm]{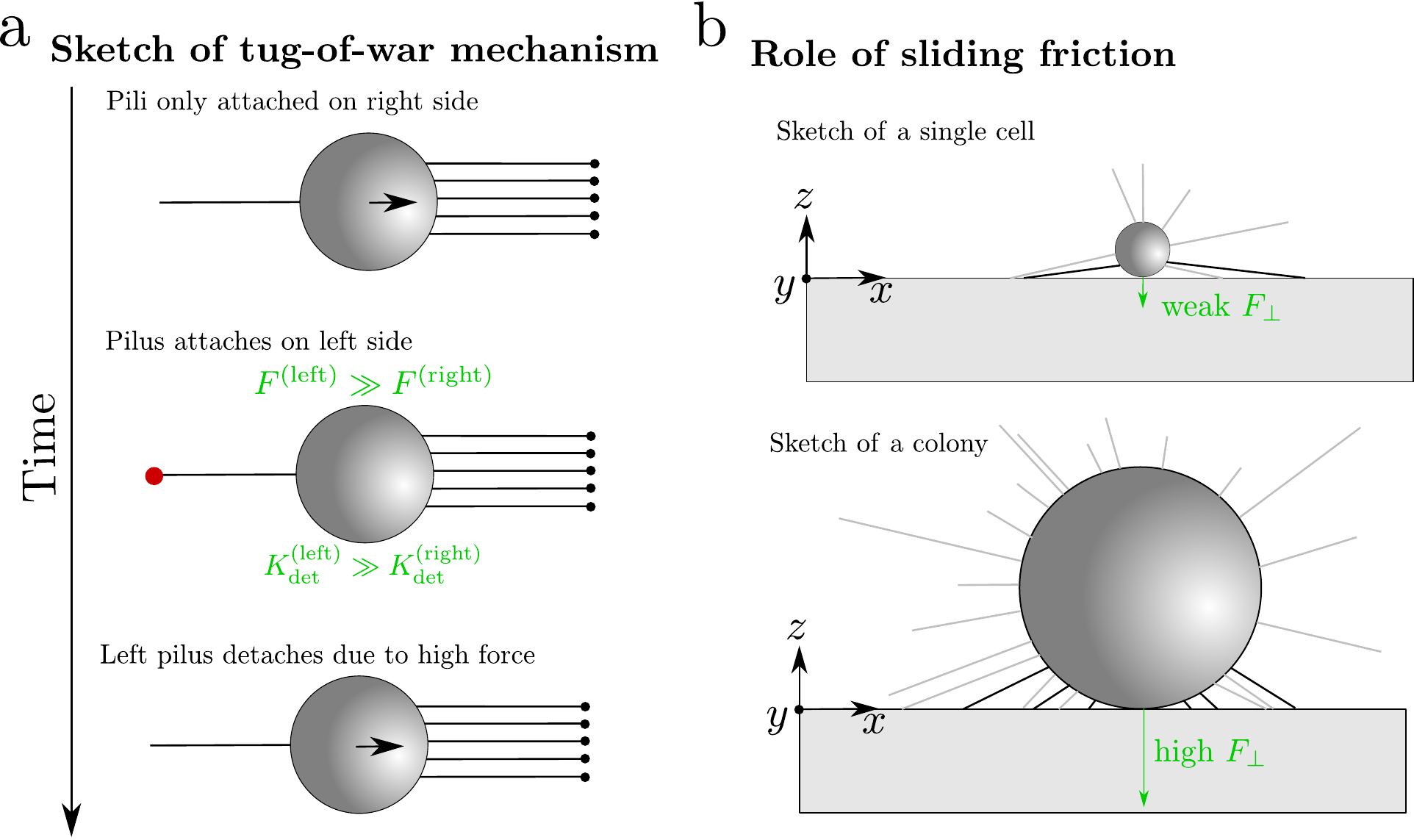}%
		\caption{\label{figmechanisms} Sketches of the tug-of-war mechanism and the role of sliding friction. (a) Sketch of how a tug-of-war can mediate a persistent motion in a given direction. Initially, a cell has five pili attached on the right side and one free pilus at the left side. When this pilus attaches to the substrate and starts to retract, it will balance the forces of the pili pulling in the opposite direction. A larger force will be exerted on the left pilus and thus, its detachment rate will be considerably higher. Thus, the left pilus will detach more easily, and a configuration where the majority of pili are attached on one side only is stable. (b) Sketch of how the sliding friction differs for cells and colonies. The spherical cell or colony possesses free (gray) and attached (black) pili. For cells, the mean angle $\amean$, characterizing how attached pili emerge from the surface, is smaller (see Fig.~\ref{figgeometry}d). Thus, the normal forces, pointing into the substrate and increasing the sliding friction force, is smaller than those of a colony that has a larger value of $\amean$ and, thus, the normal component of the force is larger. }	
	\end{figure*}
	In this paper, we investigated how type IV pili mediate the motion of individual bacteria and cell aggregates on a substrate.
	We treat the cells as active random walkers and developed a one-dimensional stochastic model, incorporating the binding and unbinding dynamics of multiple pili, the effective role of the three-dimensional geometry, and computing how the resulting force balance affects the motility of the cells. 
	We were able to show how a tug-of-war mechanism (see Fig.~\ref{figmechanisms}a) can explain the emergence of persistent motion of individual cells, without including any directional memory effects. A possible origin of the discrepancy to previously published results~\cite{marathe:2014} is the disregard of the three-dimensional structure of the cell and their pili, and how this affects the binding and unbinding dynamics of the pili.
	In previous theoretical descriptions~\cite{morikawa:2014, marathe:2014, zaburdaev:2014, brill:2017}, pili-mediated motility was studied in two-dimensional systems. In our study, with the help of a one-dimensional model that incorporates major three-dimensional effects and a three-dimensional model, we found that the shape of cells and colonies is important to understand their motion. 
	
	We discovered that a tug-of-war mechanism provided us with an explanation of pili-mediated single cell motility, but is not capable of explaining the motility of microcolonies alone. Instead, we discovered that sliding friction between the cells and the substrate can account for the observed decreasing motility of microcolonies with increasing colony size (see Fig.~\ref{figmechanisms}b).
	We found that the contribution of the friction with the substrate is increasing with microcolony size since the mean angle~ $\amean$ and, consequently, the normal component of the pili forces are increasing, pulling the colony into the substrate. For small cells and colonies the contribution of the normal force is so small that it only has a minor quantitative effect on the motility. In the future, experimental measurements quantifying the friction between bacteria and a substrate are required to verify our hypothesis.
	
	While both theoretical models, the one-dimensional stochastic model, and the three-dimensional model,  were developed to explain the peculiar behavior of the bacterium \textit{N gonorrhoeae}, they have a more general application and can be easily applied to other bacteria, such as \textit{P. aeruginosa} and \textit{N. meningitidis}, that all use related mechanisms for substrate motion. 
	The modeling approaches presented in this paper can be also applied to study the motion of a wider range of biological system characterized by a force balance of multiple active appendages, for example, cooperative transport of food by multiple ants~\cite{gelblum:2015, feinerman:2018} or the motion of sea urchin and star fish due to multiple adhesive organs, the so-called tube feet~\cite{kerkut1954, domenici:2003,cohen:2018}.
	
	\begin{acknowledgments}
		We would like to thank Nicolas Biais and Frank J\"ulicher for fruitful discussions. W.P. kindly acknowledges support from the International Max Planck Research School for Cell, Developmental and Systems Biology (IMPRS-Celldevosys). This work was supported by the Russian Science Foundation Grant No. 16-12-10496 (V.Z.).
	\end{acknowledgments}
	
	\appendix
	
	\section{\label{sec:appendixCompModel} Description of the three-dimensional model}
	In the following, we will introduce the main features of the three-dimensional model. A related version of this model, including pili mediated cell-cell-interactions, was published previously by P\"onisch et al.~\cite{poenisch:2017}.

	\subsection{Geometry of cells and pili}
\begin{figure*}[!t]
		\includegraphics[width=11cm]{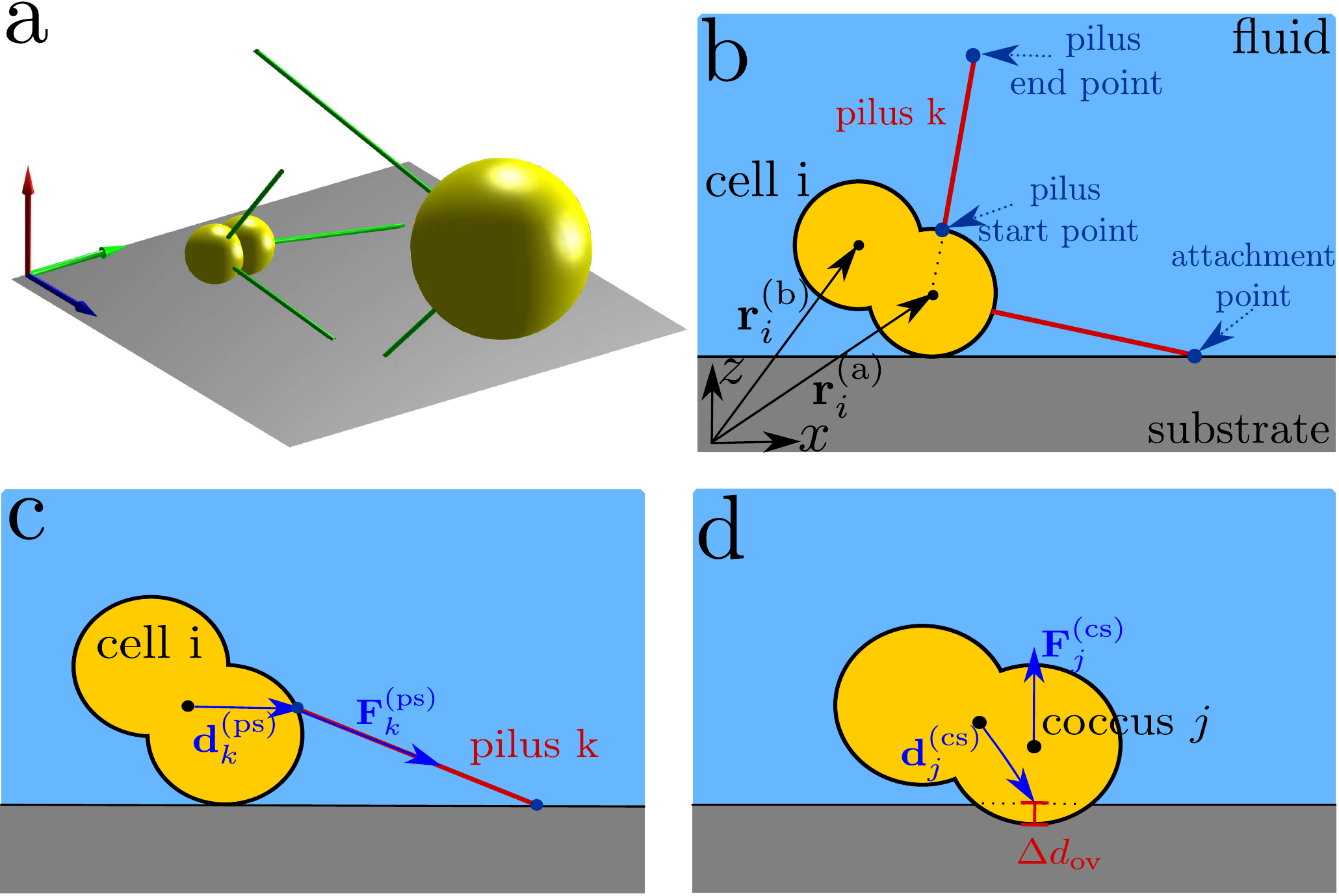}%
		\caption{\label{figcompmodel} Sketch of the three-dimensional model. (a) Three-dimensional representation of a dumbbell-shaped cell and a spherical colony with multiple pili. (b) Schematic representation of the geometry of a single cell with multiple pili sitting on the substrate, as used in the three-dimensional model. The cell consists of two cocci $a$ and $B$ with positions $\ria$ and $\rib$. The pili are characterized by a start and an end point. Pili will slide along the substrate if they are sufficiently long. Pili will attach with their tip to a substrate. (c) Sketch of pili-mediated forces due to the attachment to the substrate. Here, $\Fkps$ is the force of pilus $k$ and $\dkps$ is the distance vector pointing from the center of the cell to the pilus anchor point. (d) Excluded volume forces $\Fjsub$ of a cell overlapping with the substrate (with the overlap length $\dover$) and with the distance vector $\djsub$, pointing from the center	of the cell pointing to the point where the force is acting.}
	\end{figure*}

	We mimic the dumbbell shape of the bacteria by constructing an \textit{in silico} cell from two spheres that each have a radius $\Radius$ (see Fig.~\ref{figcompmodel}a and Fig.~\ref{figcompmodel}b). These two spheres $a$ and $b$, also called cocci, have positions $\ria$ and $\rib$ for a cell $i$ and a fixed distance $\dcocci$. The cell center is then given by 
	\begin{equation}
	\rcom = \frac{\ria + \rib}{2}.
	\end{equation}
	We will also use this model to describe the substrate motility of spherical colonies. In this case, we set ${\ria = \rib}$, which is analogous to ${\dcocci = 0}$.
	
	In the three-dimensional model, we assume that the number of pili a cell possesses, called $N$, is fixed. Pili have a mean length of ${\lchar = 0.9-1.2\ \mu \mathrm{m}}$~\cite{holz:2010,zaburdaev:2014} and, due to their semiflexible nature, we approximate pili as straight lines and specify their position by two points: their start point (also called anchor point), located at the cell surface, and their end point (see Fig.~\ref{figcompmodel}b). The distance between these two points is then called the contour length $\lcontour$ of the pilus.
	
	\subsection{Pili dynamics and pili substrate binding}
	When a new pilus is generated, it protrudes with a velocity $\vpro$ until it reaches a length resulting from an exponential length distribution with the mean length $\lchar$. Then, it will irreversibly switch to the retraction state, where a pilus retracts with a velocity $\vret$. The retraction continues until the pilus has a contour length $\lcontour = 0$. Then, the pilus is removed, and a new pilus is generated at a random position on the surface of the cell. 
	A pilus that is not interacting with the substrate will always elongate in the direction perpendicular to the cell surface. If the end point of a pilus is located within the substrate, it will not enter the substrate, but slide along its surface in the same $x$-$y$-direction as before the collision, but with the $z$-component being $0$. 
	
	Pili bind stochastically to a substrate. Here, we will assume that pili can only bind to the substrate with their tip only. The binding is modeled as a Poisson process with the binding rate $\Katt$. For this process, we assume that a pilus can only attach to the substrate if its tip sits on the substrate (see Fig.~\ref{figcompmodel}b). After the binding, the pili that were in the protrusion state will immediately start to retract~\cite{chang:2016}. 
	
	Attached pili generate pulling forces. To include such forces, we model pili as Hookean springs with the spring constant $\kspring$~\cite{biais:2010}. If the pilus $k$ is attached to the substrate, it will be stretched due to its retraction and mediates a pulling force $\Fkps$ (see Fig.~\ref{figcompmodel}c). This force is proportional to the difference between the contour length $\lcontour$ and the length of the pilus if it would not be attached, here called the free length $\lfree$. For a non-attached pilus we set contour length and the free length equal, ${\lcontour=\lfree}$. Additionally, a pilus can only generate a pulling force and no pushing force, thus, the force ${\Fkps = 0}$ if ${\lfree > \lcontour}$.  The force ${F=|\Fkps|}$ generated by the pilus $k$ affects the retraction velocity, with the stalling force $\Fstall$ (see Eq.~\eqref{eq:stalling}). The force of a pilus will also affect the unbinding from the substrate. The detachment rate is given by Eq.~\eqref{eq:detrate}.
	
	\subsection{Cell forces and motility}
	Cells are not only affected by the pili-mediated pulling forces, but also by excluded volume forces with the substrate. If a cell overlaps with a substrate, located at position $z = 0$ (see Fig.~\ref{figcompmodel}d), a repulsive force $\Fjsub$ will act from the substrate on coccus $j=a,b$, described by a Hookean spring with spring constant $\kcs$, in the normal direction of the substrate.
	The total force of the cell $i$ is given by
	\begin{equation}
	\Ftot = \sum_{\mathrm{cocci}\ j} \Fjsub + \sum_{\mathrm{pili}\ k} \Fkps.
	\end{equation}
	Additionally, cells are able to rotate. Their total torque results from
	\begin{equation}
	\Ttot = \sum_{\mathrm{cocci}\ j} \djsub \times \Fjsub + \sum_{\mathrm{pili}\ k} \dkps \times \Fkps.
	\end{equation} 
	In these two equations, $\djsub$ and $\dkpp$ describe the distance vectors pointing from the center of the cell $\rcom$ to the point at which the force is acting (see Fig.~\ref{figcompmodel}c and Fig.~\ref{figcompmodel}d).
	Finally, the total force and torque of a cell is translated into the motion and rotation of the cell $i$ by 
	\begin{equation} \label{eq:overdampedmotion}
	\frac{\mathrm{d} \rcom}{\mathrm{d} t} = \frac{\mathrm{d} \ria}{\mathrm{d} t} = \frac{\mathrm{d} \rib}{\mathrm{d} t} = \mobtrans \Ftot
	\end{equation}
	and
	\begin{equation}
	\wtot = \mobrotat \Ttot.  
	\end{equation}
	Here, $\mobtrans$ and $\mobrotat$ are the translational and rotational mobilities and $\wtot$ is the angular velocity vector. We assume that the overdamped limit is valid~\cite{purcell:1977}.
	
	For a sphere dragged through a fluid in the low Reynolds-number regime, the Stokes relation states that the translational mobility $\mobtrans$ is proportional to $\Radius^{-1}$~\cite{purcell:1977}, in particular in the absence of boundaries ${\mobtrans = 1 \slash \left( 6\pi \viscosity \Radius \right)}$. 
	For a colony of size $\Radius = 10\ \mu \mathrm{m}$ and the viscosity of water, we find ${\mobtrans \approx 10\ \mu \mathrm{m} \slash (\mathrm{s \ pN})} $. For a single cell of radius ${0.7\ \mu \mathrm{m}}$ we find ${\mobtrans \approx 100\ \mu \mathrm{m} \slash (\mathrm{s \ pN})} $. 
	This mobility is higher than the one chosen for the motility of single cells in this study. This is motivated by the numerical speedup that a lower mobility of ${\mobtrans \approx 10\ \mu \mathrm{m} \slash (\mathrm{s \ pN})} $ offers, while this value acts as a lower limit for which we can neglect the role of the colony size-dependence of the hydrodynamic friction. 
	For this mobility, a force of $0.2\ \mathrm{pN}$ causes a motion of the cell or colony with the characteristic retraction velocity $\vret$ of the pili. Importantly, this force is significantly smaller than the detachment forces (see Eq.~\eqref{eq:detrate}). 
	In the three-dimensional model, we will consider the size-dependence of the rotational mobility and set $\mobrotat = 20 \slash \Radius \ (\mathrm{s \ pN})^{-1}$.
	We will neglect hydrodynamic interactions of the spherical colony with the substrate~\cite{goldman:1967}. Such interactions of colonies close to a substrate would decrease the translational mobility maximally to $20-30\ \%$~\cite{goldman:1967}. 
	
	In this study, we will also investigate the role of sliding friction for the motion of colonies on a substrate~\cite{persson:2000}. To account for this process, we compute the normal force of all attached pili of a cell $i$ pulling the cell into the substrate, called $\Finorm$. Then, the force $\slidcoeff \Finorm$ is opposing the $x$-$y$-component of the total force $\Ftot$. If the absolute value of $\Ftot$ is smaller than the absolute value of $\Finorm$, then the cell will not be able to move in this direction. 
	
	\subsection{Parameters and simulation details}
	\begin{table}[tb]
		\caption{\label{table_Suppparametersets}%
			Parameters of the three-dimensional model. These parameters are valid for the simulations of a single cell. For the simulations of spherical colonies, we use different geometries, affecting the radius~$\Radius$ and the cocci distance~$\dcocci$, and different rotational mobilities~$\mobrotat$, accounting for the colony size dependence of the friction for spherical aggregate.
		}
		\begin{ruledtabular}
			\begin{tabular}{lcr}
				\textrm{ Name}&
				\textrm{Value}&
				\textrm{Ref}\\
				\colrule
				Radius $\Radius$\ $[\mu \mathrm{m}]$  & 0.5 & \cite{zaburdaev:2014}  \\ 
				Coccus distance $\dcocci$\  $[\mu \mathrm{m}]$ & 0.6 & \cite{zaburdaev:2014} \\ 
				Excl. volume constant\ $\kcs$ $[\mathrm{pN}\slash\mu\mathrm{m}]$  & $8 \times 10^4$ & -  \\ 
				Translational mobility\ $\mobtrans$ $[\mu\mathrm{m} \slash (\mathrm{s\ pN})]$ & 10 & \cite{marathe:2014}  \\ 
				Rotational mobility\ $\mobrotat$ $[1 \slash (\mu\mathrm{m} \ \mathrm{s\ pN})]$ & 20 & -  \\ 
				Protrusion velocity\ $\vpro$ $[\mu\mathrm{m} \slash \mathrm{s}]$ & 2 & \cite{maier:2013}  \\ 
				Retraction velocity\ $\vret$ $[\mu\mathrm{m} \slash \mathrm{s}]$ & 2 & \cite{maier:2013}  \\ 
				Mean pili length\ $\lchar$ $[\mu\mathrm{m}]$ & 1.5 & \cite{holz:2010, zaburdaev:2014}  \\ 
				Pili spring constant\ $\kspring$ $[\mathrm{pN} \slash \mu\mathrm{m}]$ & 2000 & \cite{biais:2010}  \\ 
				Pili stalling force\ $\Fstall$ $[\mathrm{pN}]$ & 180 & \cite{maier:2013, marathe:2014}  \\ 
			\end{tabular}
		\end{ruledtabular}
	\end{table}
	
	All parameters were either motivated by experimental measurements or only played a minor role for the outcome of the simulations (see Table~\ref{table_Suppparametersets}).
	
	The simulations were performed on the local computing cluster of the MPI-PKS, consisting of x86-64 GNU/Linux systems. All machines possess Intel Xeon processors with a clock rate of 2.2 to 3.0 GHz and have between 2 to 4 CPUs. The code was written in C++ and parallelized on CPU by using the library OpenMP. We used the GCC-compiler (version 4.8.5) and were running the simulations on up to 8 cores in parallel. We used an Euler algorithm to solve the equations of motion with a time step ${\Delta t = 1 \times 10^{-6}\ \mathrm{s}}$. Higher order schemes offer comparable results, but they do not increase the computation speed. For the random distribution of pili start points on the spherical cell surface we used the GNU Scientific Library (GSL).

	\section{\label{sec:appendixGeometricModel} Geometric model for the estimation of the effective attachment rate and the mean position of pili attachment}
	
	First, we calculate the probability that a pilus, growing from a random point on the surface of a spherical colony of radius $\Radius$, will attach to the substrate with its tip. The anchor point of the pilus is defined by the inclination angle ${\SurfAngle \in [0,\pi]}$ (see Fig.~2a). A pilus needs to reach at least a length 
	\begin{equation}\label{eq:lmin}
	\lmin (\Radius,\SurfAngle) = -R \left( 1 + \frac{1}{\cos \SurfAngle}\right)
	\end{equation}
	to be able to bind to the substrate. The pili have an exponential length distribution
	\begin{equation}\label{eq:pililengthdist}
	\plength(\length) = \frac{1}{\lchar} \exp \left( -\frac{\length}{\lchar}\right),
	\end{equation}
	with the characteristic length $\lchar$~\cite{holz:2010}. The binding kinetics are described as a Poisson process with the binding rate $\Katt$. The probability distribution of attachment times $t$ is then given by
	\begin{equation}\label{eq:pdfattachment}
	\patt (t) = \Katt \exp \left( -\Katt t\right).
	\end{equation}
	A pilus of length $\length$, given by the distribution $\plength(\length)$, protrudes until it reaches the minimal pili length $\lmin$, intersects with the substrate, and continues its protrusion for the length ${\length - \lmin}$. Then, it starts to retract over the same distance until it becomes shorter than $\lmin$ and loses contact to the substrate. The protrusion and retraction velocity were shown to have comparable values and are assumed to be given by the same value $\velocity$~\cite{maier:2013}. Thus, a pilus of length $\length$ is in contact with the substrate for a total time of 
	\begin{equation}
	\tcontact(\Radius,\SurfAngle)  = \frac{2(\length - \lmin(\Radius,\SurfAngle))}{ \velocity},
	\end{equation}
	where the factor $2$ is a result of the cycle of protrusion and retraction. The probability that a pilus of length $\length$ will attach within the time interval ${[0, \tcontact]}$ is given by
	\begin{eqnarray}
	\Pattmin(\Radius,l, \SurfAngle) &=& \int_{0}^{\tcontact(\Radius,\SurfAngle)}\mathrm{d}t\ \patt (t) \nonumber \\ &=& 1 - \exp\left[ - \frac{2 (\length - \lmin(\Radius,\SurfAngle))}{\latt}\right],
	\end{eqnarray}  
	where we introduced the attachment length ${\latt=\velocity \slash \Katt}$. Now, we can determine the attachment probability for random pili lengths, but constant angle $\SurfAngle$, given by
	\begin{eqnarray}
	\Pattangle(\Radius,\SurfAngle) &=& \int_{\lmin(R,\SurfAngle)}^{\infty}\mathrm{d}\length\ \plength(\length) \Pattmin(\Radius,\length,\SurfAngle) \nonumber \\ &=& \Pgamma \exp \left[ -\frac{\Radius}{\lchar} \left( 1 + \frac{1}{\cos \theta} \right) \right], \label{eq:Pattangle}
	\end{eqnarray}
	with the prefactor ${\Pgamma = 2 \lchar \slash \left( 2 \lchar + \latt \right)}$. The exponential term describes the probability of the pili to reach the substrate and is independent of the attachment rate. Only the prefactor $\Pgamma$ is affected by the stochastic binding dynamics. Pili that do not point towards the substrate have the attachment probability ${\Pattangle (\SurfAngle \leq \pi \slash 2 ) = 0}$. 
	We can interpret the attachment probability of pili $\Pattangle$ as a colony surface density of the probabilities. From this equation, we want to calculate the total number of pili that intersect and bind to the substrate. The total number of attached pili results from the integration of this density over the colony surface, 
	\begin{eqnarray}
	\Natt(R) &=& \surfdens \int_\surface \mathrm{d}\surface \ \Pattangle(\Radius,\SurfAngle) \nonumber \\ &=& \Ntot(R) \Pgamma \nonumber \\ &&\times \left[ 1 + \frac{\Radius}{\lchar} \exp \left( \frac{\Radius}{\lchar} \right) \mathrm{Ei} \left(-\frac{\Radius}{\lchar}\right)\right],
	\label{eq:attachednumber}
	\end{eqnarray}
	where ${\surface = \mathrm{d}\phi \mathrm{d}\SurfAngle R^2 \sin \SurfAngle} $ describes the colony surface in spherical coordinates (radius $\Radius$, polar angle $\SurfAngle$, azimuthal angle $\phi$) and where we use the exponential integral function $\mathrm{Ei}(x)$, the pili surface density $\surfdens$ and the total pili number of the spherical colony, 
	\begin{equation}
	\Ntot(R)=2 \pi \surfdens \Radius^2.
	\end{equation} 
	Note that for the computation of $\Ntot$ we only consider the half of the spherical colony that is exposed toward the substrate (${0 \leq z < \Radius}$). Importantly, Eq.~\eqref{eq:attachednumber} is a monotonically increasing function. Thus, the larger a colony, the more pili are able to attach to a substrate. 
	
	Finally, the probability to attach to a substrate for a single pilus that emerges from a microcolony is given by
	\begin{eqnarray}\label{eq:AttRatio2}
	\Patt(R) &=& \frac{\Natt(R)}{\Ntot(R)} \nonumber \\ &=& \Pgamma \left[ 1 + \frac{\Radius}{\lchar} \exp \left( \frac{\Radius}{\lchar} \right) \mathrm{Ei} \left(-\frac{\Radius}{\lchar}\right)\right].
	\end{eqnarray} 
	For increasing colony size, the probability that an arbitrary pilus is able to attach to the substrate is decreasing.
	
	The attachment probability $\Katt$ for a pilus tip in vicinity of the substrate is required to estimate the effective attachment rate $\Keff$ of a pilus. To estimate this rate, we first determine the time after which a pilus attaches to the substrate, assuming that this pilus will indeed attach during its lifetime (the time between its production and destruction). If a pilus emerges from the colony surface at an angle $\SurfAngle$, it first needs to protrude over a length $\lmin$ (see Eq.~\eqref{eq:lmin}), taking a time ${\tmin = \lmin \slash \velocity}$. Afterwards, it attaches to the substrate with a rate $\Katt$. This corresponds to an average time
	\begin{eqnarray}
	\AttTime(\Radius,\SurfAngle) &=& \tmin(\Radius,\SurfAngle) + \frac{1}{\Katt} \nonumber \\ &=& -\frac{R}{\velocity} \left( 1 + \frac{1}{\cos \theta} \right) + \frac{1}{\Katt} 
	\end{eqnarray}
	and the mean attachment rate averaged over the complete colony surface, is then given by
	\begin{equation}
	\Keffatt(R) = \frac{\int_\surface \mathrm{d}\surface \ \frac{1}{\AttTime(\Radius,\SurfAngle)}}{\int_\surface \mathrm{d}\surface \ } = \velocity \frac{\latt - \Radius + \Radius \ln \left( \frac{\Radius}{\latt}\right)}{\left( \latt - \Radius\right)^2}.
	\end{equation}
	Not all pili are able to reach or to bind to the substrate. To take this fact into account, we now multiply this rate by the attachment probability $\Patt(R)$ of an arbitrary pilus (see Eq.~\eqref{eq:AttRatio2}):
	\begin{equation}\label{eq:effattrate2}
	\Keff(R) = \Gamma(R) \Keffatt(R).
	\end{equation}
	As a next step, we determine the distance between the cell anchor point and the substrate attachment point of a pilus $\Lmean$, as shown in Fig.~2b. 
	To this end we consider the two contributions to the full length of an attached pilus:
	\begin{enumerate}
		\item{First, a pilus needs to reach the length $\lmin$ to be in the vicinity of the substrate.}
		\item{Afterward, a pilus will protrude until it reaches a length ${x=\lturn}$ resulting from its exponential length distribution $\plength(x)$, and then starts to retract.
			The probability distribution of attachment during the protrusion is given by taking the distribution of attachment times (see Eq.~\eqref{eq:pdfattachment}) and also given by
			\begin{equation}
			\patt(x) = \frac{1}{\latt} \exp \left( - \frac{x}{\latt}\right)
			\end{equation}
			and folding at the length $\lturn$, describing the switching from the protrusion state to the retraction state. Then, the probability density function of a pilus binding at position $x$ is given by
			\begin{equation}\label{eq:bindingpointdist}
			\pturn(\lturn,x) = \frac{\exp \left(-\frac{x}{\latt} \right) + \exp \left(-\frac{2\lturn - x}{\latt} \right)}{\latt - \latt \exp \left( -\frac{2\lturn}{\latt} \right)}. 
			\end{equation} 
			Due to the stochasticity of the binding (with the rate $\Katt$), not all pili will bind, and the probability that a pilus will bind to the substrate is given by 
			\begin{eqnarray}\label{eq:attprobturn}
			\Pturn(\lturn) &=& \int_{0}^{2\lturn \slash \velocity} \mathrm{d}t\ \patt  \nonumber \\ &=& 1 - \exp \left( -\frac{2\lturn}{\latt}\right).
			\end{eqnarray}
			We now consider the exponential length distribution of the pili $\plength$ (see Eq.~\eqref{eq:pililengthdist}) with the characteristic length $\lturn$, instead of pili with a fixed length $\lturn$. To compute the average point of attachment, we need to multiply this probability density function by the probability of attachment to
			the surface for a given length, $\Pturn$ (see Eq.~\eqref{eq:attprobturn}), to consider the fact that not all pili will attach to the substrate. We use this weight function in order to determine the mean of the binding point distribution $\pturn$ (see Eq.~\eqref{eq:bindingpointdist}):
			\begin{eqnarray}
			\ptwo(x) &=& \frac{\int_{x}^{\infty}\mathrm{d}\lturn'\ \pturn(\lturn',x) \Pturn(\lturn') \plength(\lturn')}{\int_{0}^{\infty}\mathrm{d}\lturn'\ \Pturn(\lturn') \plength(\lturn')} \nonumber \\
			&=&\frac{1}{\lzero} \exp \left( - \frac{x}{\lzero} \right), \label{eq:pdflzero}
			\end{eqnarray}
			with the length scale $\lzero = \latt \lchar \slash \left( \latt + \lchar \right)$. This is the probability distribution of the length contribution of pili that have already reached the substrate, ${\length > \lmin}$. 
			For large attachment rates ${\latt \gg \lchar}$ this length converges to $\lzero \rightarrow \lchar$, so that, if the attachment rate is low, the length scale of the pilus attachment corresponds to its length distribution. For the case ${\latt \ll \lchar}$, the length converges to $\lzero \rightarrow 0$. In this case, pili will immediately attach to the substrate when they reach it and this length scale is negligible.
		}
	\end{enumerate}
	From these length scales and their distribution (see Eqs.~\eqref{eq:lmin} and~\eqref{eq:pdflzero}), we can define the probability distribution of lengths of attached pili as a function of the pili position, given by the angle $\SurfAngle$, and the radius of the cell or colony $\Radius$:
	\begin{equation}
	\ptotallength(\Radius,\length,\SurfAngle) = 
	\begin{cases} 
	0 &\mbox{if } \length \leq \lmin(\Radius,\SurfAngle) \\
	\ptwo(\length - \lmin(\Radius,\SurfAngle)) & \mbox{if }\length > \lmin(\Radius,\SurfAngle)
	\end{cases} \label{eq:pdftotallength} .
	\end{equation}
	For a pilus length $\length$ and an angle $\theta$, the length $\Lproj$, projected on the substrate
	is then given by
	\begin{equation}
	\Lproj(\Radius,\length,\SurfAngle) = \sqrt{\length^2 - \height(\Radius,\SurfAngle)^2},
	\end{equation}
	with the height $\height$ of the anchor point (see Fig.~2b)
	\begin{equation}
	\height(\Radius,\SurfAngle) = \Radius + \Radius \cos \SurfAngle.
	\end{equation}
	To compute the mean projected length $\Lmean$, we define a new weight function from the probability distribution of the attachment points (see Eq.~\eqref{eq:pdftotallength}) and multiply it by the probability for pili emerging from a point, defined by the radius $\Radius$ and the angle $\SurfAngle$, (see Eq.~\eqref{eq:Pattangle}).
	The mean projected length $\Lmean$ then results from
	\small
	\begin{equation}
	\Lmean(R) = \frac{\int \mathrm{d}\surface\ \int_{\lmin(R,\SurfAngle)}^{\infty} \mathrm{d}l'\ \Lproj(\Radius,\length', \SurfAngle) \ptotallength(\Radius,\length',\SurfAngle) \Pattangle(\Radius,\SurfAngle)}
	{\int \mathrm{d}\surface\ \int_{0}^{\infty} \mathrm{d}l'\ \ptotallength(\Radius,\length',\SurfAngle) \Pattangle(\Radius,\SurfAngle)},
	\label{eq:meanlength}
	\end{equation}
	\normalsize
	where $\surface$ describes the surface of the lower half of the spherical colony. Importantly, here we have ${\int_{0}^{\infty}\mathrm{d}l\ \ptotallength(\Radius,\length,\SurfAngle) = 1}$. The integral is solved numerically. We see that for both parameter sets, the length has a maximal value for a specific colony size $\Radius$. Additionally, the length can exceed the characteristic pili length, particularly for the \textit{slow} parameter set. For the \textit{fast} parameter set, the average distances are smaller than those for the \textit{slow} parameter set, because the attachment rate is considerably higher and pili will immediately attach to the substrate when being in its vicinity. For the \textit{slow} parameter set, the pili are still able to elongate before attachment.
	
	In a similar manner, we can calculate the mean angle $\amean$ with which a pilus is leaving the colony (see Fig.~2b). For a pilus of length $\length$, protruding from the surface of the cell at a point given by the angle $\SurfAngle$ (see Fig.~2a), the angle is given by 
	\begin{equation}
	\aangle(\Radius,\SurfAngle,\length) = \arccos \left[ - \frac{\Radius}{\length} \left(1 + \cos \SurfAngle \right) \right].
	\end{equation}
	The mean angle is then given by
	\small
	\begin{equation}\label{eq:meanangle}
	\amean(R) = \frac{\int \mathrm{d}\surface\ \int_{\lmin(\Radius,\SurfAngle)}^{\infty} \mathrm{d}l'\ \aangle(\Radius,\SurfAngle,\length') \ptotallength(\Radius,\length',\SurfAngle) \Pattangle(\Radius,\SurfAngle)}{\int \mathrm{d}\surface\ \int_{0}^{\infty} \mathrm{d}l'\ \ptotallength(\Radius,\length',\SurfAngle) \Pattangle(\Radius,\SurfAngle)},
	\end{equation}
	\normalsize
	and again solved numerically. We find that for increasing colony size, the value of the mean angle is increasing. Thus, the force acting on the substrate in normal direction will also increase.
	
	\section{\label{sec:piliforces} Model of stationary pili forces}
	
	\begin{figure}[!tb]
		\centering
		\includegraphics[width=8cm]{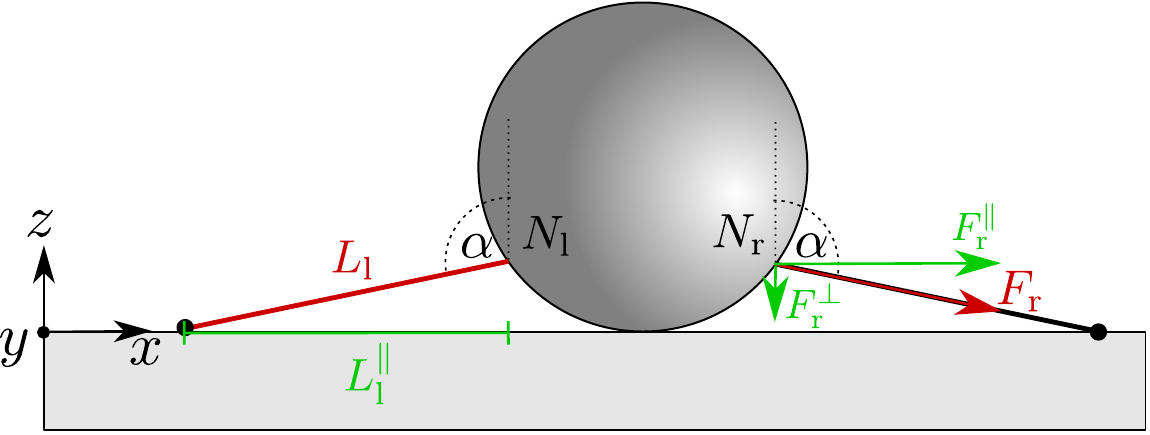}
		\caption{\label{figoneDmodel} Estimating the pulling force of pili in a simplified system. A spherical cell or colony is located on top of a substrate. It possesses $\Nleft$ pili on the left side and $\Nright$ pili on the right side. The pili have lengths $\Lleft$ and $\lright$ and are attached to the substrate. They emerge from the surface of the spherical aggregate by an angle $\aangle$ and pull with forces $\Fleft$ and $\Fright$. For those forces, we can compute the normal components, $\Fleftperp$ and $\Frightperp$, and the tangential components, $\Fleftpara$ and $\Frightpara$. The same we can do for the lengths, to compute $\Lleftperp$, $\Lleftpara$, $\Lrightperp$ and $\Lrightpara$.}	
	\end{figure}
	
	In this appendix, we will develop a model that enables us to study how pili build up a pulling force and how these forces are translated into the motion of the corresponding cell/microcolony.
	The system we investigate (see Fig.~\ref{figoneDmodel}) consists of a single spherical cell or colony that possesses $\Nleft$ attached pili on its left side and $\Nright$ attached pili on its right side. The pili point from the surface of the colony into a direction is described by the angle $\aangle$ (see Fig.~\ref{figoneDmodel}). They are modeled as Hookean springs with a fixed spring constant $\kspring$ and the contour lengths $\Lleft$ and $\Lright$, describing the length between the anchor point of the pilus and its attachment point at the left or right side. If the same pilus would not be attached to the wall, its length, called $\lleft$ or $\lright$,  could be different from the contour length. The force that a pilus produces is given by
	\begin{equation}
	\Fleft = \kspring \left( \Lleft - \lleft \right)\ \mathrm{or} \ \Fright = \kspring \left( \Lright - \lright \right). 
	\label{eq:leftrightforce}
	\end{equation}
	The components of these forces acting parallel to the substrate are given by
	\begin{equation}
	\Fleftpara = \Fleft \sin \aangle \ 
	\end{equation}
	and
	\begin{equation}
	\Frightpara = \Fright \sin \aangle
	\end{equation}
	and the components acting into the normal direction of the substrate are given by
	\begin{equation}
	\Fleftperp  = - \Fleft \cos \aangle 
	\end{equation}
	and
	\begin{equation}
	\Frightperp = - \Fright \cos \aangle.
	\end{equation}
	Here, we chose the value of the normal component of the forces such that they are positive if the pilus points into the substrate.
	
	In this model, we assume that a cell or colony can only move tangentially to the substrate. In this case, the sum of tangential forces $\Fleftpara$ and $\Fleftpara$ are those that mediate the motion of the cell. Additionally, we include sliding friction with the friction coefficient $\slidcoeff$. Here, we will make the assumption ${\Nleft \leq \Nright}$. All other cases can be calculated in an equivalent manner due to symmetry. Then, the total force that pulls the cell or colony in the right direction is given by 
	\begin{eqnarray}
	\Ftotlr &=& \max \left[0,\ \left(-\Nleft \Fleftpara + \Nright \Frightpara \right) - \slidcoeff \left( \Nleft \Fleftperp + \Nright \Frightperp \right) \right]  \nonumber \\
	&=& \max \left[ 0,\  \Nleft \Fleft \left( - \sin \aangle + \slidcoeff \cos \aangle \right) 
	\right.\nonumber \\
	&&+ \left. \Nright \Fright \left( \sin \aangle + \slidcoeff \cos \aangle \right) \right].
	\end{eqnarray}
	We connect the change of the position $x_\mathrm{c}$ of the center of the cell or colony to the length of the pili:
	\begin{eqnarray}
	\Lleftdiffpara &=& \Lleftdiff \sin \aangle =  \dot{x}_\mathrm{c}, \\ \Lrightdiffpara &=& \Lrightdiff \sin \aangle =  -\dot{x}_\mathrm{c},
	\end{eqnarray}  
	where ${\Lleftpara = \Lleft \sin \aangle}$ and ${\Lrightpara = \Lright \sin \aangle}$
	are the pili contours lengths, projected on the substrate. Then,
	\begin{eqnarray}
	\Lleftdiff  &=&  \frac{\dot{x}_\mathrm{c}}{\sin \aangle}, \\ 
	\Lrightdiff &=& -\frac{\dot{x}_\mathrm{c}}{\sin \aangle}.
	\end{eqnarray} 
	We can also express the velocity of the cell by considering the translational mobility $\mobtrans$ (see Eq.~\eqref{eq:overdampedmotion}):
	\begin{equation}
	\dot{x}_\mathrm{c} = \mobtrans \Ftotlr.
	\end{equation}
	By considering the stalling of pili, we can define the force-dependence of the change of the free length of pili (see Eq.~\eqref{eq:stalling}) and get:
	\begin{equation}
	\lleftdiff = -\max \left[ 0,\ \vret \left( 1 - \frac{\Fleft}{\Fstall}\right) \right]
	\end{equation}
	and
	\begin{equation} \lrightdiff = -\max \left[ 0,\ \vret \left( 1 - \frac{\Fright}{\Fstall}\right) \right],
	\end{equation}
	with the retraction velocity $\vret$ and the stalling force $\Fstall$. By calculating the time derivative of Eq.~\eqref{eq:leftrightforce} and combining all these equations, we can write
	\begin{eqnarray}\label{eq:Forceleft2}
	\Fleftdiff &=& \kspring \mobtrans \max \left[ 0,\ \Nleft \Fleft \left( - 1 + \slidcoeff \cot \aangle \right) \right. \nonumber \\
	&&+ \left.  \Nright \Fright \left( 1 + \slidcoeff \cot \aangle \right) \right] \nonumber \\
	&&+ \kspring \vret \max \left[ 0,\   1 - \frac{\Fleft}{\Fstall} \right] 
	\end{eqnarray}
	and 
	\begin{eqnarray}\label{eq:Forceright}
	\Frightdiff &=& - \kspring \mobtrans \max \left[ 0,\ \Nleft \Fleft \left( - 1 + \slidcoeff \cot \aangle \right) \right. \nonumber \\
	&&+ \left.  \Nright \Fright \left( 1 + \slidcoeff \cot \aangle \right) \right] \nonumber \\
	&&+ \kspring \vret \max \left[ 0,\   1 - \frac{\Fright}{\Fstall} \right] 
	\end{eqnarray}
	Together with the initial conditions
	\begin{equation}
	\Fleft (0) = 0\ \mathrm{and} \ \Fright (0) = 0
	\end{equation}
	we can solve this system of ordinary differential equations. Therefore, we assume that the length of the pilus is very long, such that we can neglect cases in which it would reach the free length ${\lleft < 0}$ or ${\lright < 0}$. Additionally, we will assume that the angle $\aangle$ is constant.
	
	In the following, we will compute the  forces for the stationary state different configurations ${(\Nleft,\Nright)}$. Here, we will neglect the relaxation dynamics by assuming that pili are infinitely stiff. Then, we can solve the system of equations
	\begin{eqnarray}
	0&=&\Fleftdiff (\Fleft, \Fright), \\
	0&=&\Frightdiff (\Fleft, \Fright).
	\end{eqnarray} 
	
	\subsection{Pili only on one side}
	First, pili are only attached to the right side so that we need to solve 
	\begin{eqnarray}\label{eq:Forceleft2}
	0 &=& - \kspring \mobtrans \max \left[ 0,\  \Nright \Fright \left( 1 + \slidcoeff \cot \aangle \right) \right] \nonumber \\ &&+ \kspring \vret \max \left[ 0,\   1 - \frac{\Fright}{\Fstall} \right]. 
	\end{eqnarray}
	The solution is given by
	\begin{equation}
	\Fright=	\begin{cases}
	\frac{\Fstall}{1 + \frac{\Fstall \mobtrans \Nright}{\vret} \left( 1 + \slidcoeff \cot \aangle \right)}& \text{if } \left( 1 + \slidcoeff \cot \aangle \right) \geq 0\\
	\Fstall              & \text{if } \left( 1 + \slidcoeff \cot \aangle \right) < 0
	\end{cases}.
	\end{equation}
	The velocity of the cell is then given by 
	\begin{eqnarray}
	\velocity &=&	\begin{cases}
	\mobtrans \Nright \Frightpara & \text{if } \left( 1 + \slidcoeff \cot \aangle \right) \geq 0\\
	0              & \text{if } \left( 1 + \slidcoeff \cot \aangle \right) < 0
	\end{cases} \nonumber \\ &=& \begin{cases}
	\mobtrans \Nright \Fright \sin \aangle& \text{if } \left( 1 + \slidcoeff \cot \aangle \right) \geq 0\\
	0              & \text{if } \left( 1 + \slidcoeff \cot \aangle \right) < 0
	\end{cases}.
	\end{eqnarray}
	
	\subsection{Similar number of pili on two sides}
	In the case of equal number $N$ of pili on both sides, the forces on both sides are equal ${\Fleft = \Fright = F}$ and we get
	\begin{eqnarray}
	0 &=& \kspring \mobtrans \max \left[ 0,\ 2 N F \slidcoeff \cot \aangle \right]
	\nonumber \\ && + \kspring \vret \max \left[ 0,\   1 - \frac{F}{\Fstall} \right].
	\end{eqnarray}
	Due to ${\cot \aangle < 0}$, this simplifies to 
	\begin{equation}
	0 = \kspring \vret \max \left[ 0,\   1 - \frac{F}{\Fstall} \right].
	\end{equation}
	Thus, the pili on both sides will reach the stalling force:
	\begin{equation}
	\Fleft(\Nleft=N,\Nright=N) = \Fstall  
	\end{equation} 
	and
	\begin{equation}
	\Fright(\Nleft=N,\Nright=N) = \Fstall.  
	\end{equation} 
	Here, it is important to note that the forces are no longer dependent on the number of pili. Additionally, the velocity of the cell is $v=0$.
	
	\subsection{Different number of pili on two sides}
	For an unequal case of pili binding on both sides of the cell, we need to solve
	\small
	\begin{eqnarray}
	0 &=& \kspring \mobtrans \max \left[ 0,\ \Nleft \Fleft \left( - 1 + \slidcoeff \cot \aangle \right) 
	+  \Nright \Fright \left( 1 + \slidcoeff \cot \aangle \right) \right] \nonumber \\
	&&+\kspring \vret \max \left[ 0,\   1 - \frac{\Fleft}{\Fstall} \right] 
	\end{eqnarray}
	\normalsize
	and 
	\small
	\begin{eqnarray}
	0 &=& - \kspring \mobtrans \max \left[ 0,\ \Nleft \Fleft \left( - 1 + \slidcoeff \cot \aangle \right) 
	+  \Nright \Fright \left( 1 + \slidcoeff \cot \aangle \right) \right] \nonumber \\
	&&+\kspring \vret \max \left[ 0,\   1 - \frac{\Fright}{\Fstall} \right].
	\end{eqnarray}
	\normalsize
	Importantly, these equations are correct for the case ${\Nright 
		> \Nleft}$.
	For the steady state we assume that the cell or colony is trapped between the pili of both sides and thus, none of the pili continue to retract. Thus, we can state that ${\Fleft \geq \Fstall}$ and ${\Fright \geq \Fstall}$. Then, the two equations simplify to 
	\begin{equation}
	0 = \max \left[ 0,\ \Nleft \Fleft \left( - 1 + \slidcoeff \cot \aangle \right) 
	+  \Nright \Fright \left( 1 + \slidcoeff \cot \aangle \right) \right].
	\end{equation}
	This is fulfilled, if
	\begin{equation}\label{eq:diffforcesleft}
	\Fleft \geq \frac{\Nright}{\Nleft} \Fright \frac{1 + \slidcoeff \cot \aangle}{1 - \slidcoeff \cot \aangle}.
	\end{equation}
	When we start from the initial condition ${\Fleft(0) = 0}$ and ${\Fright(0) =0}$ and assume that for ${\Nleft < \Nright}$ we always have ${\Fleft(t) \geq \Fright(t)}$, it is clear that the dynamics only end if
	\begin{equation}
	\Fright = \Fstall.
	\end{equation}
	Additionally, there is no trajectory for which $\Fleft$ could exceed the lower limit given by Eq.~\eqref{eq:diffforcesleft}, unless it is lower than the stalling force. Thus, we can write
	\begin{equation}
	\Fleft = \max \left[ \Fstall,\ \Fstall \frac{\Nright}{\Nleft} \frac{1 + \slidcoeff \cot \aangle}{1 - \slidcoeff \cot \aangle} \right].
	\end{equation}  
	If pili are attached to both sides of the cell or colony, the velocity $\velocity$ is set to zero.
	
	\section{\label{sec:MFPT} Mean first passage time of attached pili on one side of a cell}
	For this model, we assume that a cell has pili attached to one side only. In total, it possesses $N$ pili. If a cell possesses $n$ attached pili, the rate for binding to a substrate is given by ${\Katt (N - n)}$ with the single pilus attachment rate $\Katt$. The detachment rate is given by ${\Kdet n}$ with the single pilus detachment rate~$\Kdet$ and the rate to move over the attached pili is given by ${n \velocity \slash l}$ with the mean pili length~$l$ and the cell velocity~$\velocity$. 
	
	Initially, a cell has $n_0$ pili attached, with
	\begin{equation}
	n_0 = \frac{\Katt }{\Katt + \Kdet + \frac{v}{l}} N.
	\end{equation}
	The Smoluchowski equation for  the probability $p(n,t)$ for this random walk is given by
	\begin{equation}
	\frac{\partial p(n,t)}{\partial t} = -\frac{\partial}{\partial n} \left[ A(n) p(n,t) \right] + \frac{1}{2} \frac{\partial^2}{\partial n^2} \left[ B(n) p(n,t) \right],
	\end{equation}
	with 
	\begin{equation}
	A(n) = \Katt (N - n) - \Kdet n - \frac{\velocity}{l} n
	\end{equation}
	and 
	\begin{equation}
	B(n) = \Katt (N - n) + \Kdet n + \frac{\velocity}{l} n.
	\end{equation}
	From~\cite{vankampen:1995} it follows for the mean first passage time $T$ to hit the point where $0$ pili are attached
	\begin{equation} \label{eq:mfpt}
	T = \int_{0}^{n_0}\mathrm{d}y\ \exp \left[\Phi \left( y\right) \right] \int_{y}^{N}\mathrm{d}x\ \exp \left[-\Phi \left( x\right) \right] \frac{2}{B(x)}
	\end{equation}
	with
	\begin{eqnarray}
	\Phi(x) &=& -\int_{0}^{x} \mathrm{d}y\ \frac{2 A(y)}{B(y)} \\ &=& - \frac{2 \left[\Katt^2 - \left(\Kdet + \frac{v}{l}\right)^2 \right] x  }{\left( \Katt - \left(\Kdet + \frac{v}{l}\right) \right)^2} \nonumber \\
	&-& \frac{ 4 \Katt \left(\Kdet + \frac{v}{l}\right) N \ln \left(1 + \frac{x}{N} \frac{\left(\Kdet + \frac{v}{l}\right) - \Katt}{\Katt} \right) }{\left( \Katt - \left(\Kdet + \frac{v}{l}\right) \right)^2}. \nonumber
	\end{eqnarray}
	\begin{figure}[!tb]
		\centering
		\includegraphics[width=7cm]{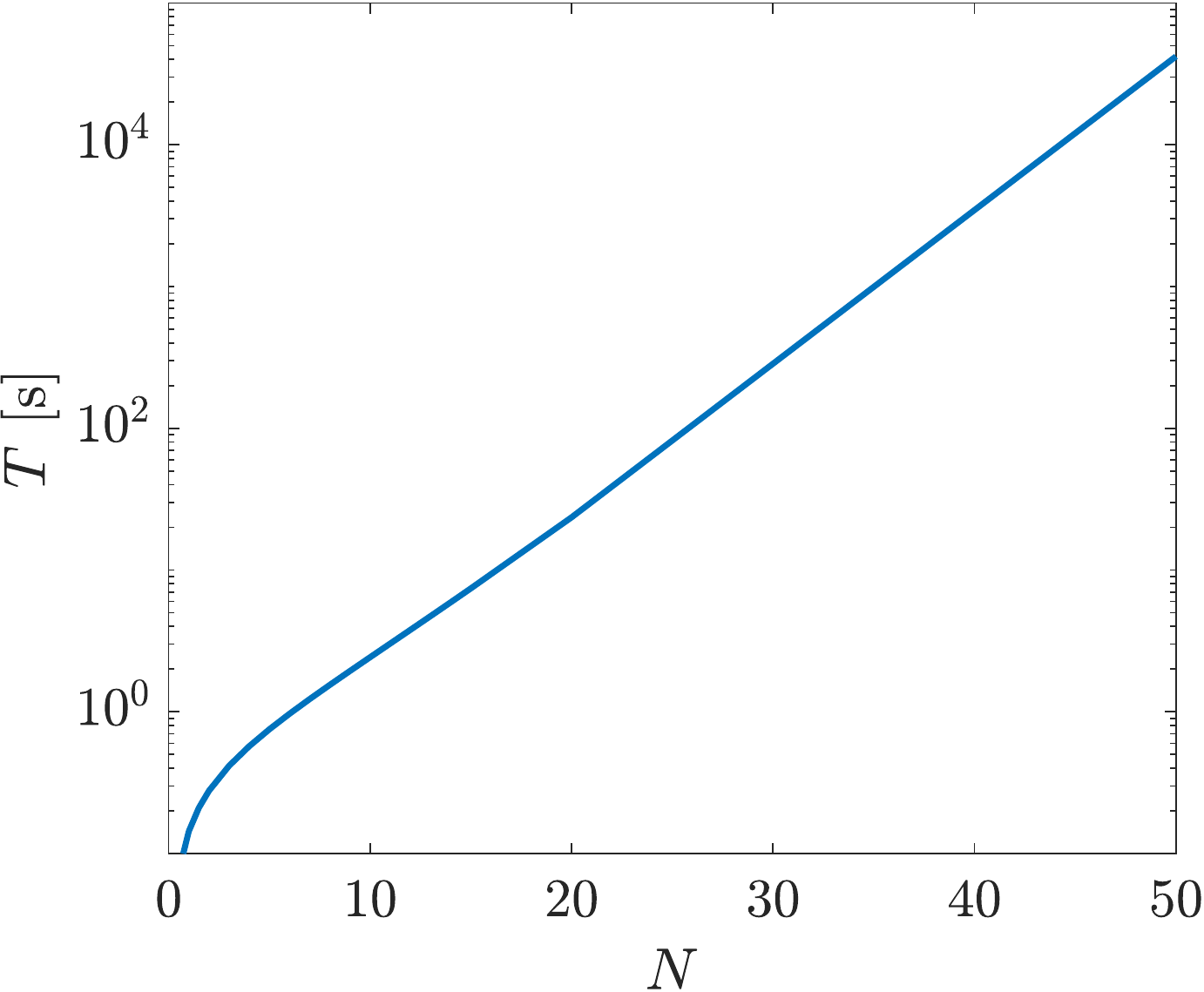}
		\caption{\label{fig:MFPT} Mean first passage time for leaving the state of pili attached to one side only by the detachment of all pili. Here, we only consider the \textit{fast} parameter set with ${\Katt = 1.95\ \mathrm{Hz}}$ and ${\Kdet = 1.12\ \mathrm{Hz}}$. Here, the detachment rate results from Table~\ref{table_parameters} by neglecting any pulling forces. Additionally, the cell velocity is given by ${\velocity = 2\ \mathrm{\mu m \slash s}}$ and the mean pili length is given by ${\latt = 0.59\ \mathrm{\mu m}}$.
		}
	\end{figure}
	We numerically solve Eq.~\eqref{eq:mfpt} and find that it is increasing exponentially for increasing pili number $N$, as shown in Fig.~\ref{fig:MFPT}.
	
	\section{\label{sec:Map1Dto2D} Mapping a one-dimensional random walk to two dimensions}
\begin{figure}[!tb]
		\centering
		\includegraphics[width=5cm]{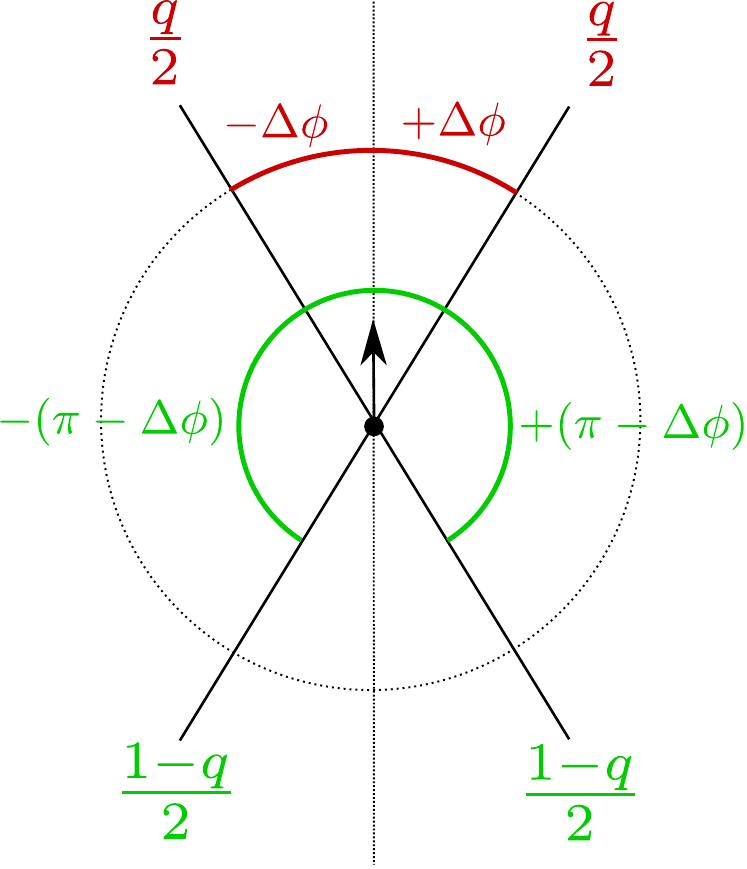}
		\caption{\label{fig:RandomWalk}  Sketch of the change of angles for the random walker. The walker changes with probability $q \slash 2$ its direction with the angle $\ood$ or $-\ood$ and with the probability ${(1-q0) \slash 2}$ in the direction ${\pi - \ood}$ or ${-(\pi - \ood)}$, starting from its current direction (black arrow). }
	\end{figure}

	In the following, we want to connect the results of the one-dimensional stochastic model to the motion over a two-dimensional substrate. Therefore, we consider a random walk that moves with a fixed velocity $v_0$, motivated by the maximal characteristic velocity of pilus retraction. 
	For every step, the random walker changes it direction either with the angle~$\ood$ or~$-\ood$. Additionally, it switches its direction with the probability~$q$, changing its direction with the angle ${\pi - \ood}$ or ${-(\pi - \ood)}$. This stochastic process is visualized in Fig.~\ref{fig:RandomWalk}.
	The probability distribution to change the angle at the step~$i$, $P(\oo)$, is given by 
	\begin{eqnarray}
	P(\oo)&=& \frac{q}{2} \delta \left( \oo - \ood \right) + \frac{q}{2} \delta \left( \oo + \ood \right) \nonumber \\ &&+ \frac{1-q}{2} \delta \left[ \oo - (\pi - \ood) \right] \nonumber \\  &&+ \frac{1-q}{2} \delta \left[\oo + (\pi - \ood) \right]
	\label{eq:Poo}
	\end{eqnarray}
	with the Dirac delta function $\delta(x)$. We assume that the time step $t_0$ is very small, so that ${q \lesssim 1}$. Then, the direction of motion after $N$ steps is given by 
	\begin{equation}
	\phi = \sum_{i=1}^{N} \oo
	\end{equation}
	where ${N=\tau \slash t_0}$ at the time $\tau$. For a constant velocity of the walker, the velocity autocorrelation function is then given by
	\begin{eqnarray}
	\mathrm{C_v}(\tau) &=& \langle \textbf{v}(t+\tau) \textbf{v}(t)   \rangle_t \nonumber \\ &=& v_0^2\langle \textbf{e}(t+\tau) \textbf{e}(t) \rangle_t \nonumber \\ &=& v_0^2 \langle \cos \phi (\tau) \rangle 
	\end{eqnarray}
	with the velocity vector~${\textbf{v}(t) = v_0  \textbf{e}(t)}$. With the probability given in Eq.~\eqref{eq:Poo} we then find
	\begin{equation}
	\langle \cos \phi (\tau) \rangle = \int_{-\infty}^{\infty}\mathrm{d}\phi_1\ P(\phi_1) ... \int_{-\infty}^{\infty}\mathrm{d}\phi_N\ P(\phi_N)  \cos \left[ \phi \right]
	\end{equation}
	and the velocity autocorrelation function
	\begin{equation}
	\mathrm{C_v}(\tau) = v_0^2 \left[ \left( 2 q - 1 \right) \cos \ood \right]^\frac{\tau}{t_0}.
	\end{equation}
	Finally, the diffusion coefficient is given by
	\begin{eqnarray}
	D &=& \frac{1}{2} \int_{0}^{\infty} \mathrm{d}\tau'\ \mathrm{C_v}(\tau') \nonumber \\ &=& -\frac{v_0^2 t_0}{2 \left[\ln \left(2 q - 1 \right) + \ln \left( \cos \ood \right) \right]}.
	\end{eqnarray}
	
	Next, we want to connect this result to the predictions of the theoretical models describing the pili-mediated motion of single cells. We will test whether we can explain the differences of the scaling of $D(N)$ that we observe in the stochastic model and the three-dimensional model.
	In Fig.~6b, we were showing that the characteristic time $\tc$ of the motion of individual cells follows:
	\begin{equation}
	\tc \propto \exp \left( \frac{N}{N_0 }\right)
	\end{equation}
	with the pili number~$N$ and the characteristic pili number $N_0$ in the stochastic model for the \textit{fast} parameter set. We now interpret this time as the time it takes the walker to switch its direction. Then, the probability to not switch direction within the time interval $t_0$, assuming that the direction swapping can be explained by a Poisson process, is given by 
	\begin{equation}
	q = \exp\left( - \frac{t_0}{\tc}\right).
	\end{equation}
	We assume that the characteristic velocity $v_0$ is not a function of $N$, but the time $t_0$, corresponding to the time it takes until the next pilus attaches to a substrate, is as follows
	\begin{equation}
	t_0 \propto \frac{1}{N}.
	\end{equation}
	Additionally, if $N-1$ pili are attached in one direction of the cell and one further pilus attaches, the change of the angle can be approximated by
	\begin{equation}
	\ood \propto \frac{1}{N}
	\end{equation} 
	based on the observation that the number of attached pili is linearly proportional to the number of pili of a cell (see Fig. 6a and 7a).
	Then, for high pili numbers~${N \rightarrow \infty}$, the diffusion coefficient fulfills the scaling
	\begin{equation}
	D \propto N
	\end{equation}
	and is no longer exponential. If we ignore the direction switching by setting $q=1$, we also find ${D \propto N}$, if on the other side we ignore the rotational diffusion and set $\ood=0$, we find ${D \propto \mathrm{e}^{N \slash N_0}}$.

	\bibliography{Literature}
	
	\clearpage
	\setcounter{figure}{0}
	\renewcommand\thefigure{S\arabic{figure}}  
	
	\begin{figure*}
		\centering
		\includegraphics[width=12cm]{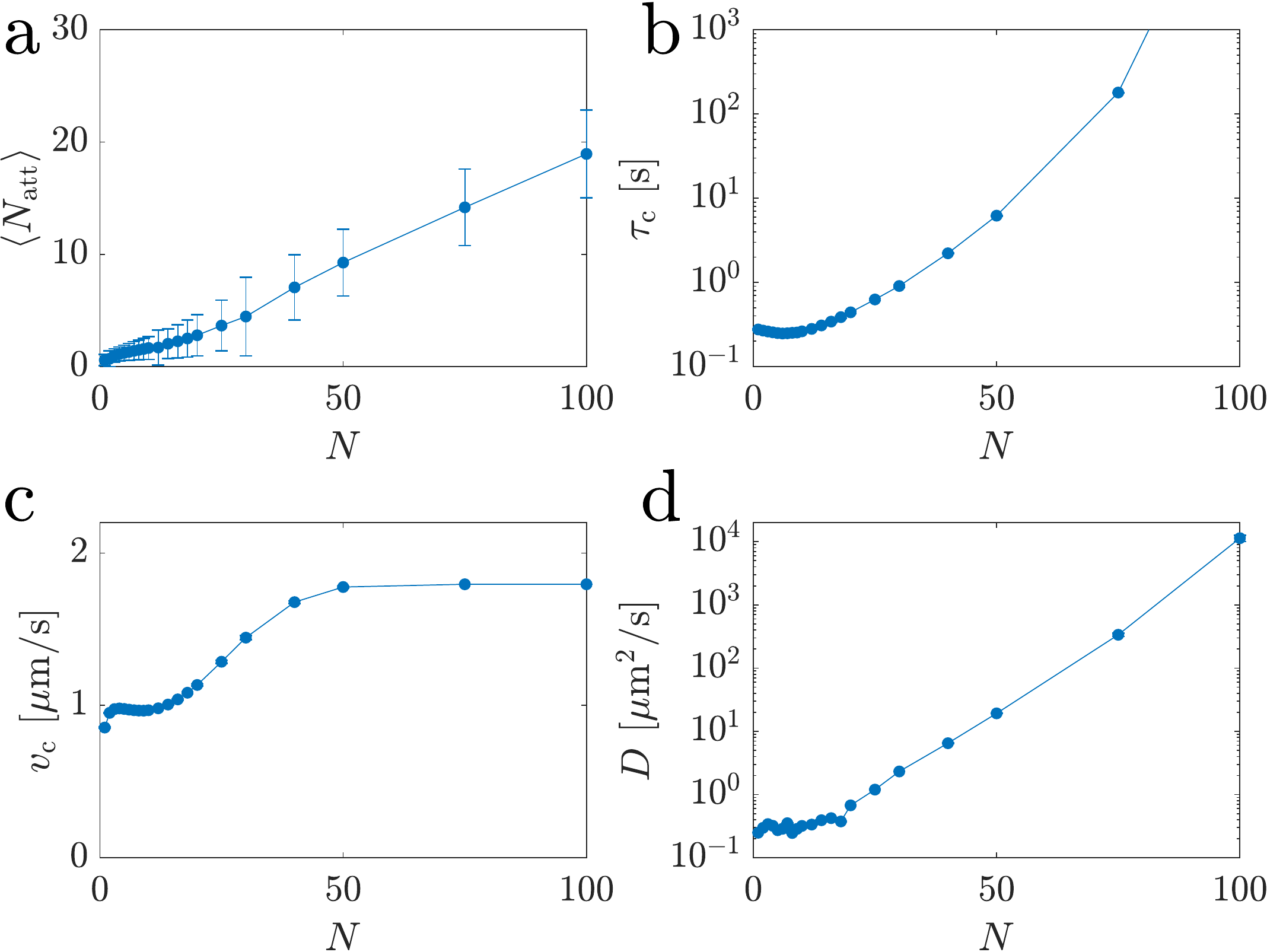}
		\caption{{\bf Pili-number dependence of an \textit{in silico} colony as calculated from the stochastic model for sliding friction coefficients ${\slidcoeff = 0.8}$.}  Data for trajectories of the \textit{fast} parameter. (a) Mean number of attached pili as a function of the total number of pili. (b) Characteristic time $\tc$ as a function of the total pili number. (c) Characteristic velocity $\vc$ as a function of the total pili number. (d) Diffusion coefficient $D$ as a function of the pili number.
			\label{fig:SS1}
		}
	\end{figure*}
	

	
	\begin{figure*}
		\centering
		\includegraphics[width=12cm]{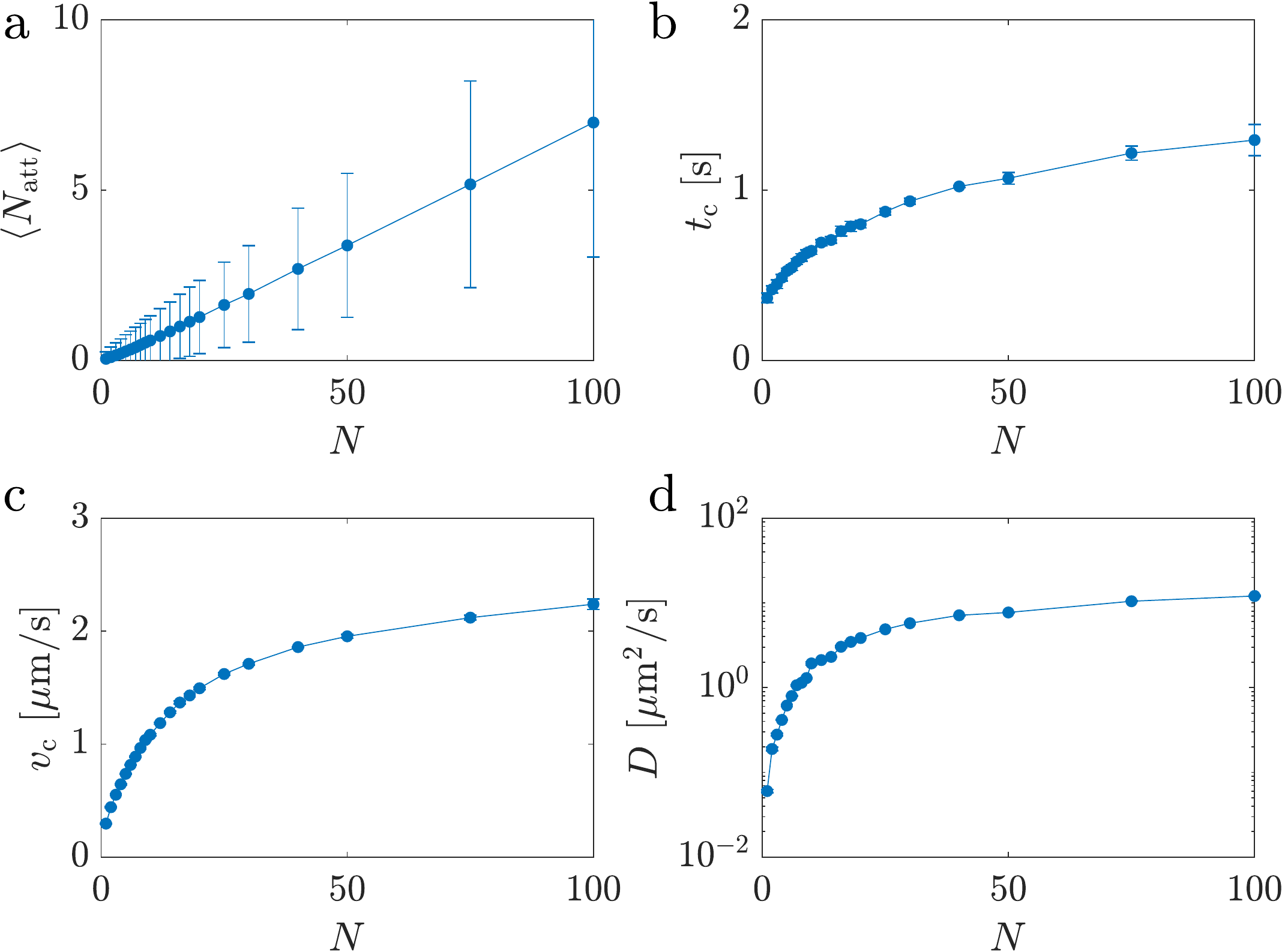}
		\caption{{\bf Pili-number dependence of an \textit{in silico} colony as calculated from the three-dimensional model for sliding friction coefficients ${\slidcoeff = 0.8}$}.  Data for trajectories of the \textit{fast} parameter. (a) Mean number of attached pili as a function of the total number of pili. (b) Characteristic time $\tc$ as a function of the total pili number. (c) Characteristic velocity $\vc$ as a function of the total pili number. (d) Diffusion coefficient $D$ as a function of the pili number.}
		\label{fig:SS2}
	\end{figure*}
	
	
	
	\begin{figure*}
		\centering
		\includegraphics[width=7cm]{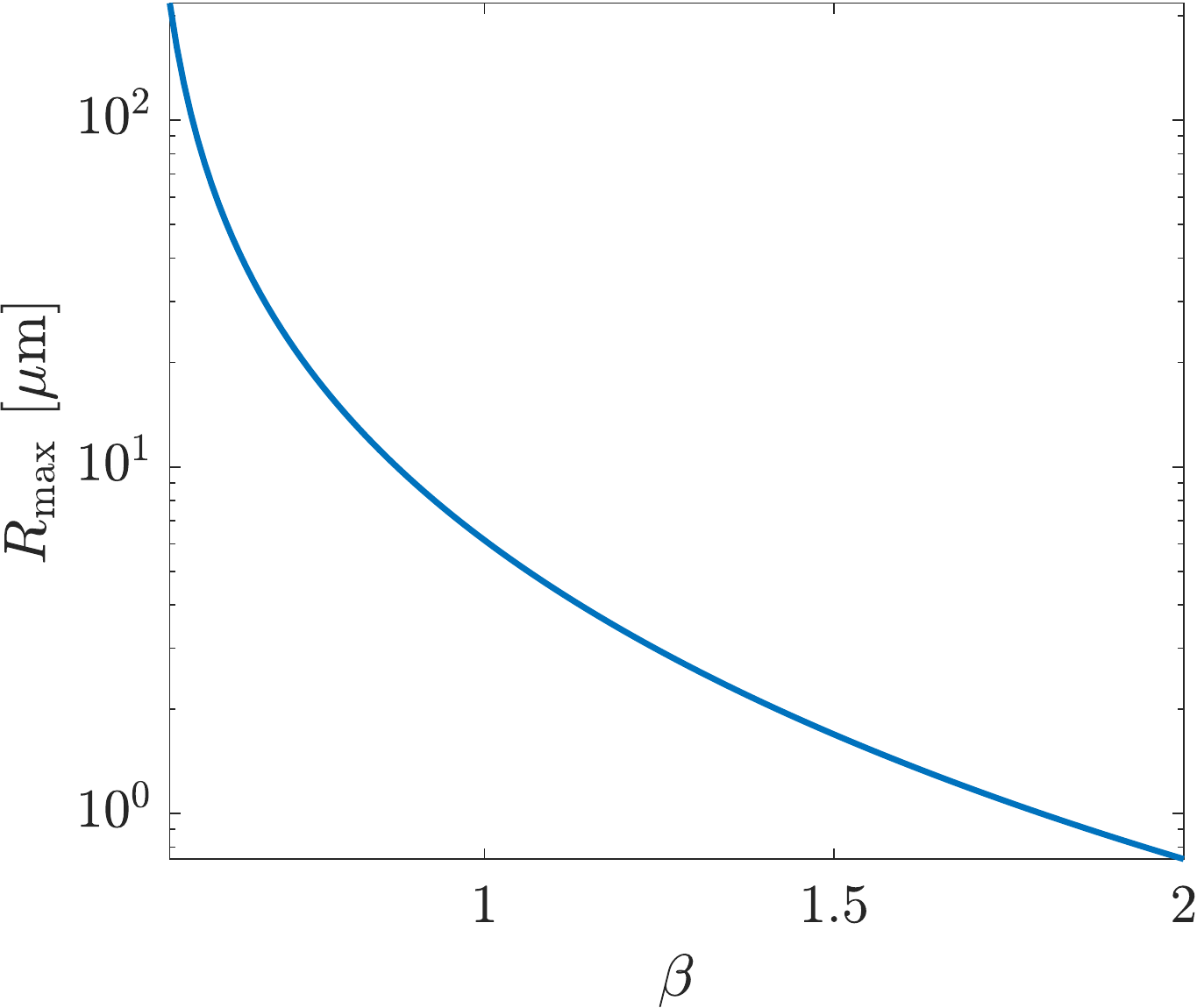}
		\caption{{\bf Maximal colony size as a function of the sliding coefficient $\slidcoeff$.}  For colonies of sizes larger than~$R_\mathrm{max}$, a colony will not be able to move for the \textit{fast} parameter set, as predicted by the calculations of Appendix~\ref{sec:appendixGeometricModel} and Appendix~\ref{sec:piliforces}.}
		\label{fig:SS3}
	\end{figure*}
	
\end{document}